%% file: 8306.tex
\documentclass{aa}
\usepackage{graphicx}
\usepackage{natbib}

\begin{document}
   \title{The close circumstellar environment of the semi-regular
   S-type star \object{$\pi ^{1}$\,Gruis}\thanks{Based on observations made with
   the Very Large Telescope Interferometer at Paranal Observatory
   under programs 077.D-0294(D/E/F)}$^{,}$\thanks{Reduced visibilities
   and differential phases are available in electronic form at the CDS
   via anonymous ftp to {\tt cdsarc.u-strasbg.fr (130.79.128.5)} or
   via {\tt http://cdsweb.u-strasbg.fr/cgi-bin/qcat?J/A+A/?/?}}}

\titlerunning{The circumstellar environment of \object{$\pi^{1}$\,Gruis}}

\authorrunning{Sacuto, S. et al.} 

   \author{S.~Sacuto\inst{1,2}, A.~Jorissen\inst{3}, P.~Cruzal\`{e}bes\inst{1}, O.~Chesneau\inst{1}, K.~Ohnaka\inst{4}, A.~Quirrenbach\inst{5}, and B.~Lopez\inst{1}}

   \offprints{S.~Sacuto}

\institute{Observatoire de la C\^{o}te d'Azur, Dpt. Gemini-CNRS-UMR 6203, Avenue Copernic, F-06130 Grasse, France
\and Department of Astronomy, University of Vienna, T\"urkenschanzstra\ss e 17, 1180 Vienna, Austria\\
\email{stephane.sacuto@obs-azur.fr ; stephane.sacuto@univie.ac.at}
\and Institut d'Astronomie et d'Astrophysique, Universit\'{e} Libre de Bruxelles, Campus Plaine C.P. 226, Boulevard du Triomphe, B-1050 Bruxelles, Belgium
\and Max-Planck-Institut f\"ur Radioastronomie, Auf dem H\"ugel 69, 53121 Bonn, Germany
\and ZAH Landessternwarte, Koenigstuhl 12, 69117 Heidelberg, Germany 
}
   \date{Received; accepted }

  \abstract 
   {}
   {We study the close circumstellar environment of the nearby S-type star
   \object{$\pi^{1}$\,Gruis} using high spatial-resolution, mid-infrared
   observations from the ESO/VLTI.}
{Spectra and visibilities were obtained with the MIDI interferometer on the VLT
Auxiliary Telescopes. The cool M5III giant \object{$\beta$\,Gruis} was used as bright primary calibrator, and a dedicated 
spectro-interferometric study was undertaken to determine its angular diameter accurately.
The MIDI measurements were fitted with the 1D
numerical radiative transfer code DUSTY to determine the dust shell
parameters of \object{$\pi^{1}$\,Gruis}. Taking into account the low spatial extension of
the model in the 8-9~$\mu$m spectral band for the smallest projected
baselines, we consider the possibility of a supplementary molecular shell.} 
   {The MIDI visibility and phase data are mostly dominated by the spherical 21\,mas (694 R$_\odot$) central star, while the extended dusty environment is over-resolved even with the shortest baselines. No obvious departure from
   spherical symmetry is found on the milliarcsecond scale. The spectro-interferometric observations are well-fitted
   by an optically thin ($\tau_{\rm dust}$$<$0.01 in the N band) dust
   shell that is located at about 14 stellar radii with a typical temperature
   of 700~K and composed of 70\% silicate and 30\% of amorphous
   alumina grains. An optically thin ($\tau_{\rm mol}$$<$0.1 in the N
   band) H$_{2}$O+SiO molecular shell extending from the
   photosphere of the star up to 4.4 stellar radii with a typical
   temperature of 1000\,K is added to the model to improve the fit in
   the 8-9~$\mu$m spectral band. We discuss the probable binary origin of asymmetries as revealed by millimetric observations.}   
{}

   \keywords{Techniques: interferometric; Techniques: high angular
                resolution; Stars: AGB and post-AGB;
                Stars: circumstellar matter; Stars: mass-loss
               }

   \maketitle
%

\section{Introduction}
\label{intro}

The class of S stars is characterized by ZrO (and sometimes LaO) bands, 
in addition to the TiO bands present in cool oxygen-rich giants of spectral
type M. The presence of lines from the unstable element Tc
(technetium) in the spectra of intrinsic S stars indicates that they
are currently evolving on the thermally-pulsing asymptotic giant
granch (TP-AGB), where internal nucleosynthesis is responsible for 
forming carbon and heavy elements (\textit{s-elements})
subsequently dredged up to the stellar surface. Being on the AGB,
intrinsic S stars experience a mass-loss allowing the formation of their
dust circumstellar environment \citep{jori98}.\\
At a distance of 153$^{+28}_{-20}$pc \citep{perr97,vane98}, 
\object{$\pi^{1}$\,Gruis} (\object{Hen\,4-202}, \object{HD\,212087}, \object{HIP\,110478}) is one of the brightest intrinsic S stars.
It is a semi-regular variable of type SRb, varying
from V=5.4 to 6.7 in 150 days. It
was chosen as an original prototype of the S star class with \object{R\,And} and
\object{R\,Cyg} \citep{merr22}, and is known to be a binary star
with a solar-type G0V companion, located at an angular distance of
2.71$\arcsec$ (unchanged for more than a century; Bonneau, priv. comm.), with a position
angle of 202.1$^{\circ}$ and an apparent visual magnitude of 10.9
\citep{feas53}. The pair is probably physical with their apparent magnitude agreeing with their spectral type for the Hipparcos distance.

With C/O ratios generally around 0.8 \citep{smit86},
photospheres of S-type stars are oxygen-rich. The situation regarding
\object{$\pi^{1}$\,Gruis} is somewhat confusing, though. According to its
position in IR color-color diagrams, \object{$\pi^{1}$\,Gruis} is classified in
the C class of intrinsic S stars \citep{jori98}. This class includes
stars with a dense oxygen-rich circumstellar envelope and with strong
chemical peculiarities and mass-loss rates ranging from a few
$10^{-7}$ to a few $10^{-5}$~M$_{\odot}$\,yr$^{-1}$. Furthermore, this
star is classified as "E" (emission silicate) by \citet{volk89}
and "SE" (silicate and oxygen-rich dust emission) by
\citet{sloa98}. The circumstellar dusty shell of \object{$\pi^{1}$\,Gruis} would
thus appear to be mainly composed of oxygen-rich elements; however, the
track in the (K-[12],[25]-[60]) color-color diagram during the
variability cycle \citep{jori98} suggests that the shell emission
could as well be associated with carbon-composite material
\citep{thom76, knap99a}.\\

The mass-loss history of the star can be traced back 21,000 years,
with the detection of a 0.28 pc extended dust shell by IRAS, assuming a 15~km\,s$^{-1}$ 
velocity of expansion \citep{youn93}. The
molecular environment was resolved in the east-west direction with a
24$\arcsec$ \citep{saha92} and a 30$\arcsec$ beam \citep{knap99a}. 
Moreover, the CO lines exhibit an unusual asymmetric
double-horned profile and a high-velocity component interpreted in the
frame of a slowly expanding disk and a fast ($v \ge 90$ km\,s$^{-1}$)
north-south bipolar outflow. \citet{chiu06} report millimeter
observations at higher spatial resolution
(2.1$\arcsec$$\times$4.2$\arcsec$ beam) that confirm this
interpretation. In addition, these observations reveal
a 200 AU cavity suggesting that the mass-loss rate of \object{$\pi^{1}$\,Gruis}
has dramatically decreased over the past 90 years. Whether the highly
structured far environment of \object{$\pi^{1}$\,Gruis} (presence of disk + jet)
is a consequence of the asymmetry of the dusty wind from the primary star or of
the presence of the far companion, or even a closer undetected one, cannot be 
addressed by the millimetric observations of \citet{chiu06} due to 
insufficient spatial resolution. 

Based on the comparative analysis of
the jets and disks present in a sample of late AGB stars (among which
\object{V\,Hya} and \object{$\pi^{1}$\,Gruis}) and of proto-planetary nebulae (including
symbiotic stars), \citet{hugg07} strongly favors the second 
possibility. The properties of the disk and bipolar outflow of
\object{$\pi^{1}$\,Gruis} appear to fit nicely in a sequence going from late
(binary) AGB stars to proto-planetary nebulae. According to the binary
scenarios discussed by \citet{hugg07}, the G0V companion could be at the origin of the launching of the bipolar outflow, despite the large separation. However, \citet{maka05} and \citet{fran07} found possible evidence for yet
another companion from the discrepancy between the Hipparcos and
Tycho-2 proper motion (the Hipparcos proper motion relies on
measurements spanning only 3~years, whereas the Tycho-2 one uses
positional data spanning more than a century). Any discrepancy between
these short-term and long-term proper motions is thus indicative of
the presence 
of an unrecognized orbital motion in addition to the true proper
motion. The discrepancy is very significant in the case of
\object{$\pi^{1}$\,Gruis}. According to the analysis of \citet{fran07}, these
so-called `$\Delta \mu$ binaries' correspond to systems with orbital
periods in the range 1500-$10^{4}$~days, much shorter thus than the
6000-year period inferred by \citet{knap99a} for the G0V companion
2.7" away.  It is therefore an open question whether a third component
might be present in the system.\\

In this paper, we present the first mid-infrared interferometric
observations of this star taken with the MIDI/VLTI instrument,
secured with auxiliary telescopes (ATs) of 1.8~m that provide a beam with an FWHM of
1.1$\arcsec$ at 10~$\mu$m. In interferometric mode, measurements of
spatially correlated flux at a high spatial resolution ($\sim$20 mas)
and multi-spectral ($\lambda$/$\Delta\lambda$$\sim$30) information on
the mid-infrared source was obtained within the interferometric
field-of-view (FOV) of 1.1$\arcsec$.  The inner diameter of the
millimetric disk ($\sim$2.6$\arcsec$) found by \citet{chiu06} is then
largely beyond the present MIDI FOV, as is the distant companion
(2.71$\arcsec$). \\ 
The outline of this paper is as follows. In Sect.~\ref{obs_cal-process}, 
we present the MIDI observations and describe the calibration 
performed with the two data-reduction software packages (MIA and EWS)
developed for the MIDI focal instrument. This
calibration required considerable effort to determine the angular diameter
of the spatially resolved calibrator M5III star \object{$\beta$\,Gruis} (described in Appendix~\ref{angular_diam_calib}). 
Section~\ref{spect-inter-model} presents our spectro-photometric and interferometric 
interpretations of the star. This was done first with an analytical radiative transfer 
model which gives some insight into the expected behavior of the
mid-infrared emission from the circumstellar envelope. Then, we used the
numerical radiative transfer code DUSTY to confirm the set of parameters found with the analytical
thin-dusty-shell model and define a chemical composition of the dust shell,
taking the position of the star on the \textit{silicate dust sequence} into account.  
Moreover, to account for the larger than expected spatial extension of the object in the 8-9~$\mu$m
spectral band, we add a molecular shell to the model. In Sect.~\ref{asymm}, we 
discuss the probable binary origin of asymmetries revealed by millimetric observations. 
Finally, we summarize and conclude in Sect.~\ref{conclusion}.

\section{Observations and calibration process}
\label{obs_cal-process}

The Very Large Telescope Interferometer (VLTI) of ESO's Paranal
Observatory was used with MIDI, the MID-infrared Interferometric
recombiner \citep{lein03}. MIDI combines the light of two telescopes
and provides spectrally resolved visibilities in the N band
atmospheric window. Guaranteed time observations (GTO) of \object{$\pi ^{1}$\,Gruis} were
conducted with the VLT ATs positioned at stations A0, E0, D0, and G0, providing baselines in
the range of 15-64 meters. Observations were made during the nights
of May 21-23-24-25-27 2006, June 19, 2006, and August 8, 2006, under
good atmospheric conditions (average seeing=0.94$\arcsec$), with the
worst seeing (=1.51$\arcsec$) during the night of May 27, 2006.
Table~\ref{journal} presents the journal of the interferometric
observations. The calibrator \object{$\beta$\,Gruis} was either observed right before
or after each science target observation.\\

\begin{table}[tbp]
\caption{Journal of observations.}
\label{journal}
\begin{minipage}[h]{10cm}
\begin{tabular}{cccccc}
\hline
\hline
{\tiny Star} & {\tiny UT date \& Time} & {\tiny $\phi^{*}$} & {\tiny Config.} & {\tiny B [m]} & {\tiny PA [deg]} \\
\hline
{\tiny \object{$\pi^{1}$\,Gruis}} & {\tiny 2006-05-21 09:09:26} & {\tiny \ldots} & {\tiny E0-G0} & 15.7 & 50 \\
{\tiny \object{$\beta$\,Gruis}} & {\tiny 2006-05-21 09:28:33} & {\tiny 0.23} & - & - & - \\
{\tiny \object{$\pi^{1}$\,Gruis}} & {\tiny 2006-05-23 09:32:27} & {\tiny \ldots} & {\tiny A0-G0} & 63.5 & 56 \\
{\tiny \object{$\beta$\,Gruis}} & {\tiny 2006-05-23 10:06:51} & {\tiny 0.28} & - & - & - \\
{\tiny \object{$\pi^{1}$\,Gruis}} & {\tiny 2006-05-24 09:27:34} & {\tiny \ldots} & {\tiny D0-G0} & 31.8 & 56 \\
{\tiny \object{$\beta$\,Gruis}} & {\tiny 2006-05-24 09:44:39} & {\tiny 0.31} & - & - & - \\
{\tiny \object{$\pi^{1}$\,Gruis}} & {\tiny 2006-05-25 08:09:13} & {\tiny \ldots} & {\tiny A0-G0} & 61.2 & 41 \\
{\tiny \object{$\beta$\,Gruis}} & {\tiny 2006-05-25 07:51:28} & {\tiny 0.34} & - & - & - \\
{\tiny \object{$\pi^{1}$\,Gruis}} & {\tiny 2006-05-27 09:20:41} & {\tiny \ldots} & {\tiny E0-G0} & 15.9 & 57 \\
{\tiny \object{$\beta$\,Gruis}} & {\tiny 2006-05-27 09:37:11} & {\tiny 0.39} & - & - & - \\
{\tiny \object{$\pi ^{1}$\,Gruis}} & {\tiny 2006-06-19 10:16:01} & {\tiny \ldots} & {\tiny D0-G0} & 30.6 & 83 \\
{\tiny \object{$\beta$\,Gruis}} & {\tiny 2006-06-19 09:50:14} & {\tiny 0.01} & - & - & - \\
{\tiny \object{$\pi ^{1}$\,Gruis}} & {\tiny 2006-08-08 08:40:43} & {\tiny \ldots} & {\tiny A0-G0} & 52.6 & 102 \\
{\tiny \object{$\beta$\,Gruis}} & {\tiny 2006-08-08 08:13:56} & {\tiny 0.36} & - & - & - \\
\hline
\end{tabular}
\end{minipage}
$*$ : the pulsation phase for \object{$\beta$\,Gruis} \citep{oter06}. This was not
possible for \object{$\pi^{1}$\,Gruis} due to the non-regular character of its lightcurve.\\
\end{table}

Chopped acquisition images were recorded (f=2Hz, 2000 frames, 4~ms per
frame, 98mas per pixel) for the fine acquisition of the target. The
acquisition filter was an N-band filter. Simultaneous
interferometric and photometric measurements were performed with the
SCI-PHOT mode of MIDI, using the prism that provides a spectral dispersion of about 30. The data-reduction software packages\footnote{\tt{http://www.mpia-hd.mpg.de/MIDISOFT/,
http://www.strw.leidenuniv.nl/$\sim$nevec/MIDI/}} MIA and EWS
\citep{jaff04} were used to reduce the spectra and visibilities
\citep{ches05}. MIA is based on the power spectrum analysis and uses a 
fast Fourier transformation (FFT) to calculate the Fourier
amplitude of the fringe packets, while EWS uses a shift-and-add
algorithm in the complex plane, averaging appropriately modified
individual exposures (dispersed channeled spectra) to obtain the
complex visibility. \\ An extensive study of the fluctuations of the
instrumental transfer function was carried out for the baseline A0-G0
during the night of 2006 May 25 (see Table~\ref{journal_2}). The rms
scatter of these calibration measurements is $\sim$0.05 between
8~$\mu$m and 13~$\mu$m, corresponding to a relative error of 10\%. This
error in SCI-PHOT mode with ATs is greater than the typical errors
encountered with UTs with the same observing mode (2-5\%). This
discrepancy is mostly due to the limited accuracy of the photometric
extraction with ATs versus UTs.\\
All visibility and differential phase data as well as all the
characteristics of the observations are available from the CDS (Centre
de Donn\'{e}es astronomiques de Strasbourg); all data products are
stored in the FITS-based, optical interferometry data exchange format
(OI-FITS) described in \citet{paul05}.

\subsection{General description}
\label{general_description}

Due to its proximity (153~pc) and its brightness (about 900 Jy at 10
$\mu$m), \object{$\pi^{1}$\,Gruis} is an ideal mid-infrared target for
interferometric measurements with the ATs. Interferometric observations provide the visibility
measurements of the source $\widetilde{V}_{\rm sou}$, which is the product
of the source calibrated visibilities $V_{\rm sou}$ with the system
response $V_{\rm sys}$, which in turn is a combination of the atmospheric
and instrumental responses. The system response is estimated using
the ratio between visibility measurements of the calibrator
$\widetilde{V}_{\rm cal}$ and its theoretical visibilities ${V}_{\rm cal}$
\citep{vanb05}. The expression of the source calibrated visibility is

\begin{equation}
\label{calibrated_visibility}
V_{\rm sou}=\frac{\widetilde{V}_{\rm sou}}{\widetilde{V}_{\rm cal}}{V_{\rm cal}}.
\end{equation}
If the calibrator is unresolved, its theoretical visibility is close to 1 and
the measured contrast of the fringes is a direct measure of the
instrumental efficiency. If the calibrator is resolved, then the values of its theoretical
spectrally dispersed visibilities need to be very accurately
determined.\\
For our observations of \object{$\pi^{1}$\,Gruis}, which has no nearby unresolved
calibrator with enough flux ($>$100 Jy at 12~$\mu$m, corresponding to
the sensitivity of MIDI with ATs in low spectral resolution and
SCI-PHOT mode), two choices were possible: either move the telescope
far from \object{$\pi^{1}$\,Gruis} to acquire a bright enough unresolved
calibration star or to observe a bright but resolved calibrator in the
vicinity of \object{$\pi^{1}$\,Gruis}. The first choice introduces a bias 
due to the air mass difference between the source observation at
a given sky location and the calibrator observation at another
location. The second choice requires an accurate knowledge of the
angular diameter of the resolved target. The second option was chosen
for optimizing the flux at the
price of a potentially large uncertainty on the angular diameter of the
calibration star (see Appendix~\ref{angular_diam_calib}). 

\subsection{Calibrated flux of the source}
\label{calib_flux}

The MIDI flux of \object{$\pi^{1}$\,Gruis} was scaled to the MARCS model
flux of \object{$\beta$\,Gruis}, 
for which we take into account the value of the
diameter from the spectrometric method (26.8$\pm$1.3 mas; see Appendix~\ref{spectro_estimation})

\begin{equation}
F_{\rm sou}=\frac{\widetilde{F}_{\rm sou}}{\widetilde{F}_{\rm cal}}{F^{\rm marcs}_{\rm cal}},
\end{equation}
where $F_{\rm sou}$ is the calibrated flux of \object{$\pi^{1}$\,Gruis},
$\widetilde{F}_{\rm sou}$ the raw flux measurements of \object{$\pi^{1}$\,Gruis},
$\widetilde{F}_{\rm cal}$ the raw flux measurements of \object{$\beta$\,Gruis},
and $F^{\rm marcs}_{\rm cal}$ the MARCS synthetic flux of \object{$\beta$\,Gruis}.

\begin{figure}[h!]
\begin{center}
\includegraphics[width=8.5cm]{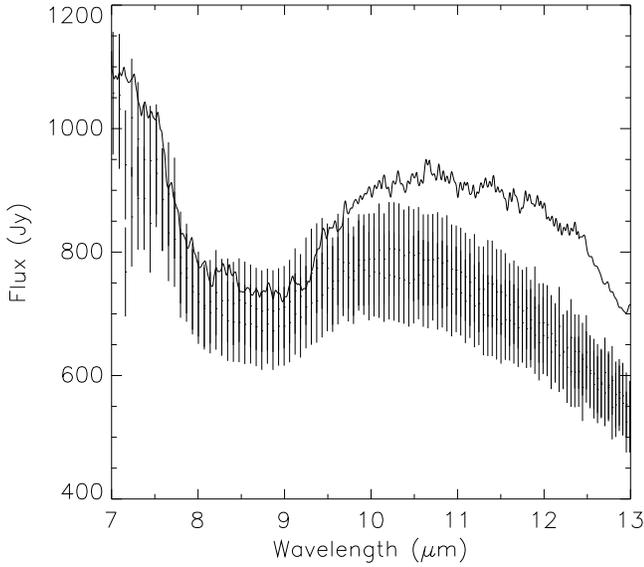}
\caption{Calibrated MIDI fluxes of \object{$\pi^{1}$\,Gruis} (vertical error
bars) as compared to the ISO/SWS spectrum (solid line) taken on 1996 Oct 26.}
\label{pigru_spectra_Jy}
\end{center}
\end{figure}

Figure~\ref{pigru_spectra_Jy} compares the single-dish MIDI spectrum with the ISO
spectrum \citep{sloa03} taken on 1996 Oct 26. Error bars are related to the uncertainty on the value of the calibrator diameter. 
The MIDI flux level is lower than the
ISO/SWS flux level. Two possibilities could account for this 
difference. The most probable one is that the beam of an AT
($\sim$1.1$\arcsec$ at 10~$\mu$m) is small compared to that of
the ISO/SWS satellite (14$\arcsec$$\times$22$\arcsec$). Therefore, because of the proximity of this star (153 pc) and the large extension of its dust envelope beyond the AT beam size, some flux might be missed. 
A second possibility that cannot be excluded is that the
level of emission has changed since the ISO observation (1996) due to
the pulsation of the star. However, this star is a semi-regular SRb pulsator with a low variability, 
for which it is difficult to assign a phase to a given observation. 
Nevertheless, we can see that the shape of the ISO/SWS and MIDI spectra is
about the same, revealing a dust feature that will be discussed in
Sect.~\ref{chem_comp}. As a consequence, we assume in the
following study that the MIDI spectrum can be complemented with the
ISO spectrum to construct the current SED of the object at longer infrared wavelengths.

\subsection{Calibrated visibilities of the source}
\label{calib_vis}

\begin{figure}[h!]
\begin{center}
\includegraphics[width=9cm]{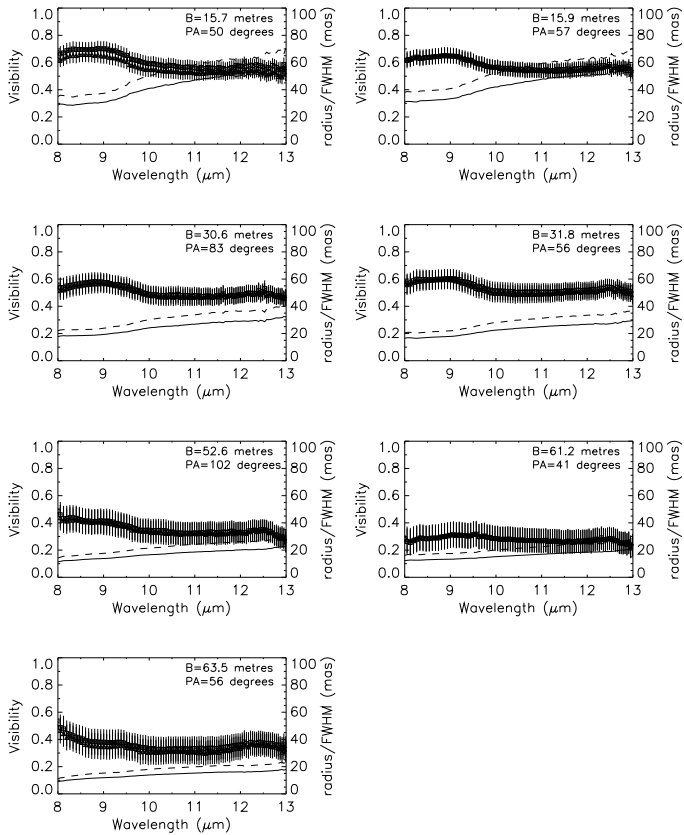}
\caption{MIDI-calibrated visibilities (error bars) with the MIA (squares) and EWS
(triangles) reduction softwares. The solid 
lines correspond to the uniform disk angular radius (in mas) and the
dashed lines to the FWHM (in mas) of the Gaussian
distribution (both to be read from the scale on the right axis),
computed from the visibility at each spectral channel.}
\label{pigru_visibility}
\end{center}
\end{figure}

In Fig.~\ref{pigru_visibility} are shown simultaneously the calibrated visibility with the MIA 
and EWS reduction softwares. The curves from the two softwares are almost indistinguishable.  
For clarity, we only show the reduction taking the spectrometric diameter of \object{$\beta$\,Gruis} into account 
(26.8$\pm$1.3~mas; see Appendix~\ref{spectro_estimation}). The error bars are the sum of the transfer function error bar of
10\% plus the error bar related to the uncertainty on the value of the
calibrator diameter (from 0.6$\%$ at low spatial frequencies to 10$\%$
at high spatial frequencies). 

The size derived from the uniform disk and the Gaussian distribution regularly increases from 8 to 13
$\mu$m (see Fig.~\ref{pigru_visibility}), indicating that an extended mid-infrared structure surrounds
the central star \citep{ohna05}. The dimension of this structure depends on the
observed wavelength, so neither of the two simple models
(monochromatic uniform-disk or Gaussian distribution) can describe the source. In
the following, we model the visibilities with radiative
transfer models taking into account the spectral dependency of the
circumstellar environment of \object{$\pi^{1}$\,Gruis}.

\begin{figure}[h!]
\begin{center}
\includegraphics[width=8.cm]{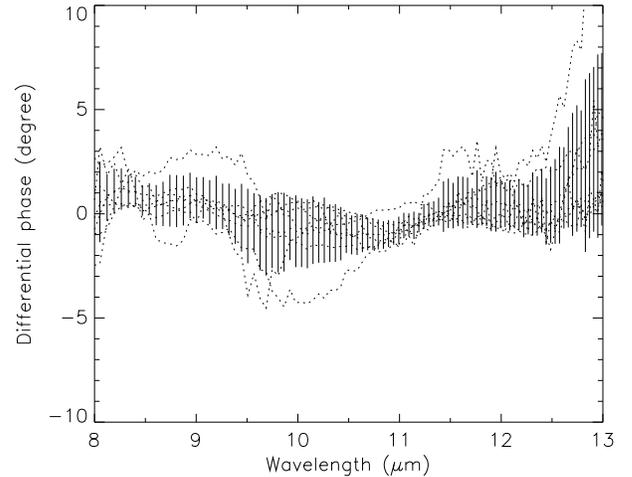}
\caption{Calibrated differential phase of \object{$\pi ^{1}$\,Gruis} evaluated
from all the determined projected baselines (dotted lines). Error bars
correspond to the rms of the differential phase averaged over all the
projected baselines.}
\label{phase_pigru}
\end{center}
\end{figure}

To check whether the object shows some asymmetries, one can
use the differential phase of \object{$\pi^{1}$\,Gruis} obtained from the EWS software. 
Figure~\ref{phase_pigru} shows the calibrated
differential phase of the object evaluated from all the determined
projected baselines. Error bars correspond to the rms of
the differential phase over all the projected baselines. The very low
rms values of the differential phase ($\pm$1.93$^{\circ}$ at 8~$\mu$m,
$\pm$1.10$^{\circ}$ at 9~$\mu$m, $\pm$1.65$^{\circ}$ at 10~$\mu$m,
$\pm$1.82$^{\circ}$ at 12.5~$\mu$m) indicates that there is no obvious
signature of an asymmetric source, such as binarity or a clumpy
environment, for the given projected baselines.

\begin{figure}[h!]
\begin{center}
\includegraphics[width=8.5cm]{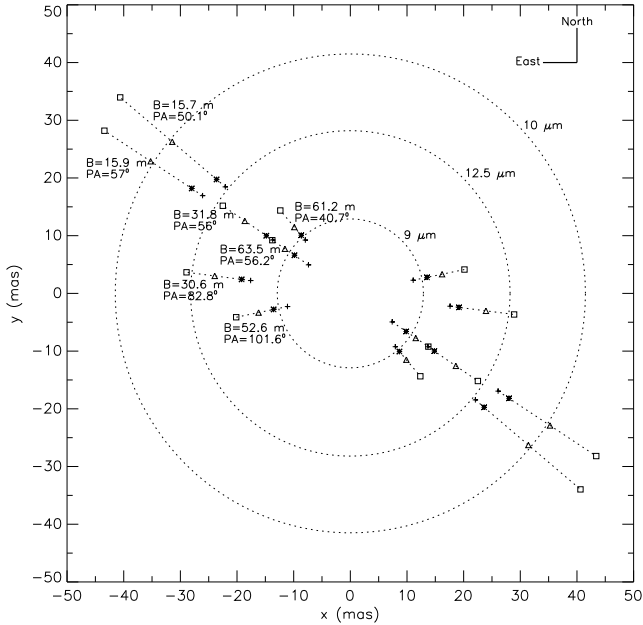}
\caption{This figure presents the dependence of the dimensions of
the equivalent uniform-disk (in mas) on the projected
baseline position
angle on the sky taken at 8 (plus), 9 (cross), 10 (triangle), and 12.5
$\mu$m (square). The three dotted circles fit the points calculated
for a given wavelength at close projected baselines
(B) for different projected baseline position angles
(PA).}
\label{symmetry_pigru}
\end{center}
\end{figure}

Furthermore, the MIDI visibility amplitudes also show no evident departure of
the object from spherical symmetry. Figure~\ref{symmetry_pigru} shows
the dependence of the dimensions of the equivalent uniform-disk on 
the projected baseline position angle on the sky at different
wavelengths. The three circles fit the points calculated for a given
wavelength at close projected baselines for
different projected baseline position angles. In addition to zero differential phase, 
these fits show that the object is nearly spherical.
It must be pointed out that for such a close, extended giant, the interferometric fringes essentially probe the
star photosphere and its close vicinity, while the extended dusty environment is over-resolved even for the 
shortest baselines.

From these considerations, spherical radiative transfer
models will be used to interpret the spectro-interferometric data of
\object{$\pi^{1}$\,Gruis}.

\section{Simultaneous fit of observed visibilities and SED}
\label{spect-inter-model}

\subsection{Photometric observations of \object{$\pi^{1}$\,Gruis}}
\label{phot_obs_pigru}

Besides the ISO/SWS spectrum (see
Fig.~\ref{pigru_spectra_Jy}), ground-based photometry is also available
for \object{$\pi^{1}$\,Gruis} (see Table~\ref{spectro-photo-pi}). The K
magnitude has already been dereddened by \citet{vane00b}. The other
magnitudes were dereddened using the Fitzpatrick absorption curve for
the M-band \citep{fitz99} and the values of \citet{riek85} for the
N-band. Van Eck's visual extinction value of A$_{\rm v}$=0.17 was 
adopted, so that the ratio of total to selective extinction at V,
3.1 on average, allowed us to correct the other magnitudes. At
wavelengths longer than 12.5 $\mu$m, the interstellar extinction
becomes negligible compared to the measurement precision of the
magnitude. Table~\ref{spectro-photo-pi} lists the basic photometric
properties of \object{$\pi^{1}$\,Gruis}. 

\begin{table}[tbp]
\caption{Broadband photometry of \object{$\pi^{1}$\,Gruis}}
\label{spectro-photo-pi}
\begin{minipage}[h]{10cm}
\begin{tabular}
[c]{ccccc} \hline\hline
Band & Wavelength & Extinction  & Dereddened & Reference \\ 
     & ($\mu$m) & coefficient & magnitude &  \\ \hline
R & 0.67 & 0.13 & 3.05  & (1) \\
I & 0.87 & 0.09 & 0.45  & - \\ \hline 
H & 1.65 & 0.03 & -1.82 & (2) \\
K & 2.2 & 0.02 & -2.07 & - \\
L & 3.6 & 0.01 & -2.50 & - \\
M & 4.8 & 0.01 & -2.26 & - \\
N & 8.4 & 0.01 & -2.89 & - \\
- & 9.7 & 0.01 & -3.25 & - \\
- & 10.5 & 0.01 & -3.51 & - \\
- & 11.2 & 0.01 & -3.58 & - \\
- & 12.5 & 0.0 & -3.47 & - \\
Q & 20.0 & 0.0 & -4.35 & - \\ \hline
U & 0.35 & 0.28 & 11.13 & (3) \\
B & 0.42 & 0.24 & 8.32 & - \\
V & 0.55 & 0.17 & 6.43 & - \\
J & 1.21 & 0.05 & -0.60 & - \\
H & 1.66 & 0.03 & -1.72 & - \\
K & 2.22 & 0.02 & -2.14 & - \\
L & 3.45 & 0.01 &  -2.57 & - \\ \hline
J & 1.25 & 0.05 & -0.71 & (4) \\
K & 2.2 & 0.02 & -2.18 & - \\
L & 3.5 & 0.01 & -2.51 & - \\
M & 4.9 & 0.01 & -2.23 & - \\
N & 12.0 & 0.0 & -3.58 & - \\ \hline
\end{tabular}
\end{minipage}
(1) \citet{john66}; (2) \citet{thom76}; \\
(3) \citet{vane00b}; (4) \citet{smit03}
\end{table}

\subsection{The stellar atmosphere of \object{$\pi ^{1}$\,Gruis}}
\label{atmos_pigru}

\begin{figure}[h!]
\begin{center}
\includegraphics[width=9cm]{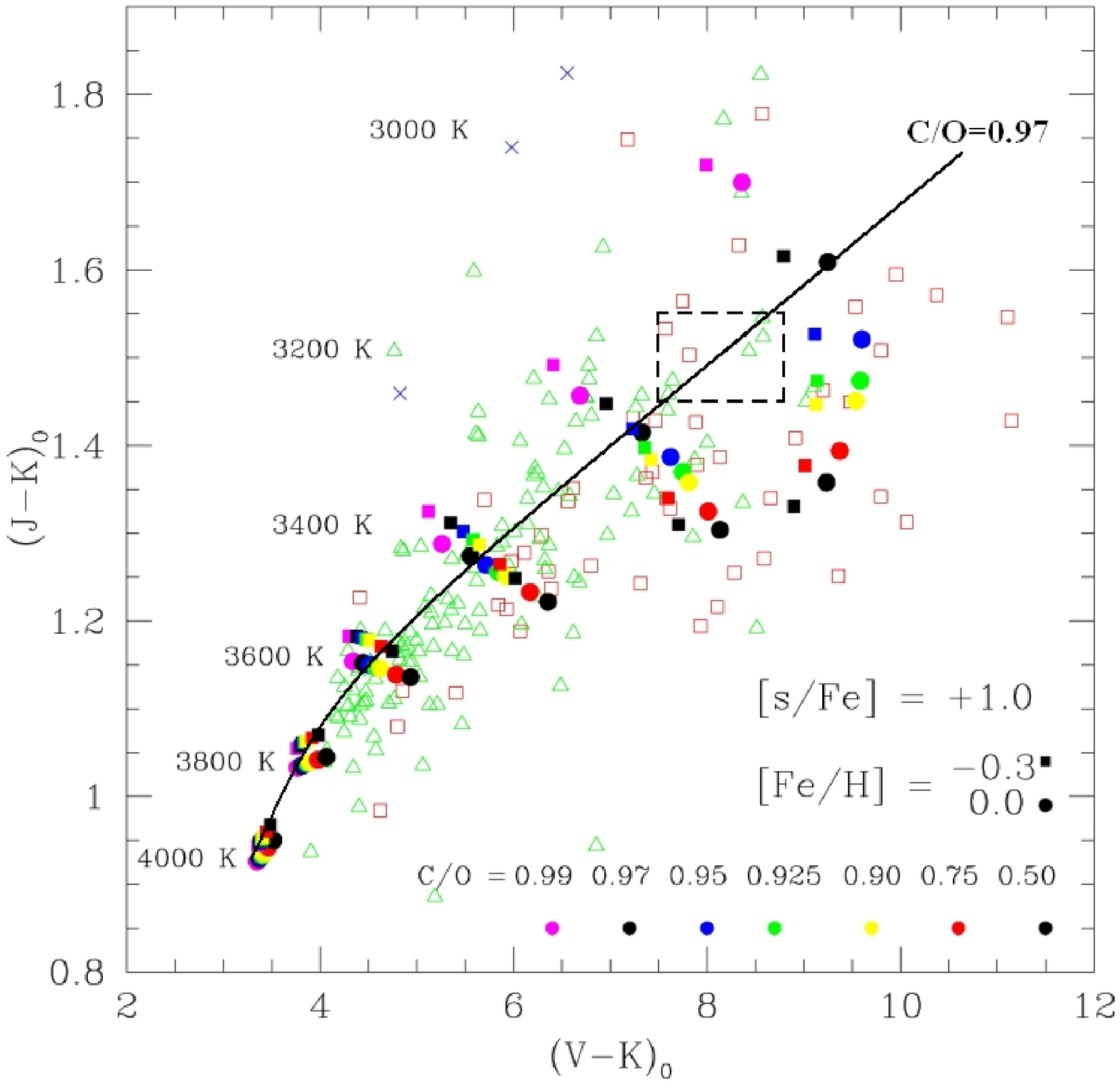}
\end{center}
\caption{Dereddened (V-K,J-K)$_{0}$ color-color diagram for S stars
(open triangles) from \citet{vane00a} and \citet{vane00b}, M giants
(open squares), and a few C stars (crosses) both from \citet{whit00},
compared to synthetic MARCS model colors (filled color symbols; \citealt{plez03}) 
of different temperatures: 4000, 3800, 3600, 3400,
3200, 3000 K. The black solid curve from lower left to upper right is
the line joining all models with C/O ratio of 0.97. The \object{$\pi^{1}$\,Gruis} 
color range is enclosed in the dashed box.}
\label{Color-Color}
\end{figure}

The realistic MARCS code (see Appendix~\ref{spectro_estimation}) was used to
determine the synthetic intensity distribution 
of the S-star \object{$\pi ^{1}$\,Gruis}. Such a distribution is far more realistic than 
a blackbody distribution due to the large number of
absorption lines in S star atmospheres.\\ The \object{$\pi ^{1}$\,Gruis} MARCS
model atmosphere is taken from a grid specifically designed for S
stars (\citet{plez03} and in preparation), since their non-solar C/O
ratio and s-process abundances impact their atmospheric
structure. Synthetic $J-K$ and $V-K$ colors were computed for
all models from this grid (Fig.~\ref{Color-Color}). These reveal that
the C/O ratio has a strong influence on the colors, especially for
low effective temperatures ($\le 3200$~K). There is, however, hardly a
one-to-one relationship between the photometric indices and the
atmospheric parameters, and several models are found to match the
$(J-K)_0$, $(H-K)_0$ and $(V-K)_0$ indices of \object{$\pi ^{1}$\,Gruis}.
All these models have $T_{\rm eff} = 3100$~K, $\log g = 0$, and [Fe/H] = 0, 
but allow some range in [s/Fe] and C/O (namely C/O = 0.97 for [s/Fe] = 0.5~dex, 
and $0.90 \le $ C/O $ \le 0.92$ for [s/Fe] = 1~dex). The situation is 
illustrated in Fig.~\ref{Color-Color}, where \object{$\pi ^{1}$\,Gruis} falls
within the dashed box. It must be noted that \citet{vant02} have performed 
a spectroscopic abundance analysis of \object{$\pi^{1}$\,Gruis}, 
though not with dedicated MARCS model atmospheres, and
found [Fe/H] $=0.05\pm0.48$ and [Zr/Ti] $= 1.27\pm0.23$ for $T_{\rm eff} = 3000$~K. 
These are in good agreement with the parameters
inferred from the MARCS models. Table~\ref{stellar-parameters-pi}
summarizes the atmospheric parameters and the most important
abundances adopted for \object{$\pi ^{1}$\,Gruis}.\\

\begin{table}[tbp]
\centering
\caption{Stellar atmospheric parameters of \object{$\pi ^{1}$\,Gruis}, all from \citet{vant02} except C/O.}
\label{stellar-parameters-pi}
\begin{tabular}
[c]{cc} \hline\hline
Parameter & Value \\ \hline
T$_{\rm eff}$ & 3000 K \\ 
log $g$ & 0.0 \\
$v_{\rm micro}$ (km\,s$^{-1}$) & 3.0 \\
$\left[Fe/H\right]$ & 0.05$\pm$0.48 \\
$\left[Zr/Ti\right]$ & 1.27$\pm$0.23 \\
C/O & 0.97 \\ \hline
\end{tabular}
\end{table}

\subsection{Analytical modeling}

Now that the MARCS intensity distribution of \object{$\pi^{1}$\,Gruis} is known,
it can be used to model the spectro-interferometric measurements of
the star.

\label{analytical_modeling}

\subsubsection{The analytical thin-dusty-shell model}
\label{analytical thin-dusty-shell model}

The general expression of the analytical thin-dusty-shell intensity
distribution was derived by \citet{cruz06}. These authors present an
analytical solution to the radiative transfer equation in spherical
geometry for an optically thin dusty shell. The model, briefly explained below, 
was compared to the well-established radiative transfer code
DUSTY used to fit the spectro-interferometric data of \object{$\pi ^{1}$\,Gruis} 
(see Sect.~\ref{dusty_code}). Using this thin-dust-shell model
is justified by the low mass-loss rate of the star (4.6$\times$10$^{-7}$ M$_{\odot}$\,yr$^{-1}$; \citealt{jori98}) in view of the 
low-contrast, mid-infrared emission features. 
This led Gail (1990) to propose that the dust formation is insufficient for triggering a dust-driven
outflow.  The advantage of this model is the straightforward way a first set of
physico-chemical parameters can be determined along with their errors. 
These output parameters are then inserted into the widely-used numerical radiative
transfer code DUSTY in order to confirm and complete the analysis of this star (see Sect.~\ref{numerical_modeling}).\\

The expression of the intensity given by the analytical
thin-dusty-shell model is described in Sect.~2.2 in \citet{cruz06}, where the central star 
is represented by the corresponding synthetic MARCS S-spectrum 
(see Sect.~\ref{atmos_pigru}). The global optical depth related to the infrared continuum excess is given by

\begin{equation}
\tau^{\rm out}_{\lambda}(\varepsilon)=\kappa^{\rm in}_{\lambda_{0}}\left(\frac{\lambda}{\lambda_{0}}\right)^{-\beta}\left(\frac{\varepsilon_{\rm in}}{\varepsilon_{\rm out}}\right)^{p} l^{\rm out}(\varepsilon),
\label{optical depth}
\end{equation}
where $\kappa^{\rm in}_{\lambda_{0}}$ is the extinction coefficient at the
inner boundary at the reference wavelength $\lambda_{0}$. The extinction 
coefficient is parametrized as a decrease with wavelength as $\lambda^{-\beta}$. 
Values of $\beta$ are usually taken between 0.2 and
2.4 depending on the chemistry, nature (amorphous or crystalline),
shape, and size of grains \citep{dupa03}. The factor $p$ corresponds to the radial
number density falling off as $r^{-p}$. The parameters $\varepsilon_{\rm in}$ and
$\varepsilon_{\rm out}$ are the angular radii of the shell inner and outer
boundaries, respectively. Finally, $l^{\rm out}(\varepsilon)$ is the total
geometrical depth of the shell.\\

The formalism described in \citet{cruz06} applies in the context of the infrared
continuum excess caused by the dust shell. However, as we can see in
Fig.~\ref{pigru_spectra_Jy}, the spectrum of \object{$\pi^{1}$\,Gruis} exhibits
a bump around 11 $\mu$m caused by the emission from oxygen-rich
dust. The optical depth of this bump can be modeled with a Gaussian
distribution \citep{groe94},

\begin{equation}
\tau_{\lambda}^{\rm bump}=A \ e^{-\left(\frac{\lambda-\lambda_{\rm c}}{\Delta\lambda}\right)^{2}},
\label{bump}
\end{equation}
where $A$ is the strength of the bump, $\lambda_{\rm c}$ the central
wavelength of the dust feature, and $\Delta\lambda$ related to the
width of the bump. Therefore, the expression for the intensity must
depend not only on the optical depth related to the infrared continuum
excess (Eq.~\ref{optical depth}) but also on the optical depth related to the mid-infrared bump (Eq.~\ref{bump}).
In the case of spherical symmetry, the visibility function
$V_{\lambda}$ measured by an interferometer is linked to the angular
distribution of the intensity of the source
$I_{\lambda}\left(\varepsilon\right)$ through the normalized Hankel
transform,

\begin{equation}
V_{\lambda}\left(\frac{b_{ij}}{\lambda}\right)=\left\vert\frac{\breve{I}_{\lambda}\left(\frac{b_{ij}}{\lambda}\right)}{\breve{I}_{\lambda}(0)}\right\vert,
\label{radial visibility}
\end{equation}
where $b_{ij}$ is the modulus of the projected baseline (on the plane
of the sky) formed by the pair of apertures $\left(i;j\right)$, and

\begin{equation}
\breve{I}_{\lambda}\left(f \right)=\int_{0}^{\infty} 2 \pi \varepsilon I_{\lambda} \left(\varepsilon \right) J_{0} \left(2 \pi f \varepsilon \right) d \varepsilon
\label{Hankel transform}
\end{equation}
is the Hankel transform of $I_{\lambda}$ at the spatial frequency $f=b_{ij}/\lambda$,
where $J_{0}$ is the zeroth-order Bessel function of the first
kind. Note that $\breve{I}_{\lambda}(0)$ corresponds to the flux
density of the source $F_{\lambda}$.

\bigskip

The aim of this model is to give a first set of physico-chemical
parameters by simultaneously fitting spectrometric and interferometric data. These
parameters are then used as input parameters in the 
numerical code DUSTY to determine the final set of parameters
characterizing the dusty circumstellar environment of \object{$\pi^{1}$\,Gruis}.

The independent parameters of the models can be determined using
the well-known iterative Levenberg-Marquardt algorithm minimizing the
$\chi^{2}$ function obtained by comparing modeled and measured visibilities,
using the weight associated with the $N$ visibility
measurements. These weights are determined by the inverse square of the visibility 
errors. The minimization process uses the analytical expressions for
the first derivatives of the visibility function with respect to each
independent parameter as extensively described in \citet{cruz06}.

\subsubsection{Spectro-interferometric results}
\label{ana_spectro_inter_result}

In this section, the \object{$\pi^{1}$\,Gruis} interferometric data are analyzed 
by means of the analytical model described in Sect.~\ref{analytical
thin-dusty-shell model}. The fit is done separately on two data 
sets, referred to as {\it DATA\_SPECTRO\_CAL} and {\it DATA\_INT\_CAL}
in the following. The former set corresponds to the \object{$\pi^{1}$\,Gruis}
MIDI data calibrated with the $26.8\pm1.3$~mas diameter of \object{$\beta$\,Gruis} derived from
the spectrometric method (see Appendix~\ref{spectro_estimation}). The latter set is calibrated
with the $28.8\pm0.6$~mas diameter of
\object{$\beta$\,Gruis} derived from the interferometric method
(see Appendix~\ref{interfero_estimation}).

Some parameters of the analytical model were fixed at pre-determined 
values or only allowed to vary within a reasonable range:

\begin{itemize}

\item The effective temperature (T$_{\rm eff}$), which was found to be 3000 to 3100~K both
from the work of \citet{vant02} and from the position of the star in the 
(J-K,V-K)$_{0}$ color-color diagram. Therefore both 3000 and 3100~K 
synthetic MARCS models were utilized for the fit.

\item A dust temperature at the inner boundary (T$_{\rm in}$) between 600 and 800 K,
in accordance with the low mass-loss rate of this star for which dust
condensation will halt at an intermediate stage of the
condensation-ladder (see Sect.~\ref{chem_comp}). 
This is called the "freeze-out" mechanism \citep{hera05, blommaert06}.

\item The central wavelength of the dust feature ($\lambda_{\rm c}$), which is fixed to 11
$\mu$m from the position of the mid-infrared bump excess (see
Fig.~\ref{pigru_spectra_Jy}).

\item Varying the density power-law coefficient ($p$) between 1 and 2.5.

\item Varying the value of the spectral index ($\beta$) between 0.2 and 1.4, in
agreement with values observed in circumstellar environments
\citep{wein89, beck90,knap93}.

\end{itemize}

\begin{figure}[h!]
\begin{center}
\includegraphics[width=8.5cm]{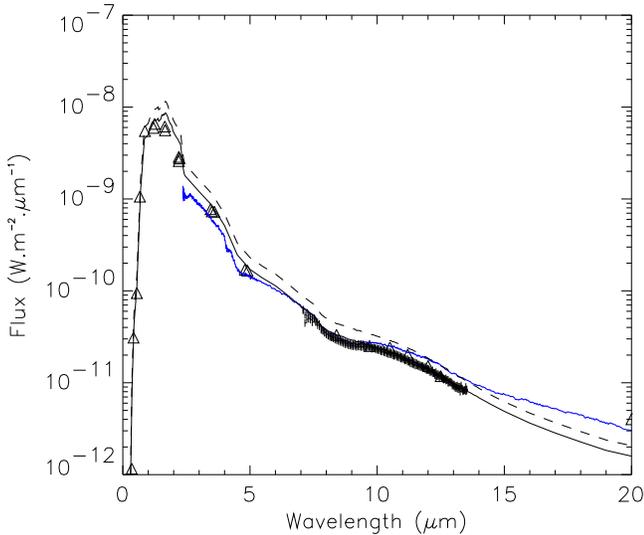} 
\end{center}
\caption{Comparison of the \object{$\pi^{1}$\,Gruis} fluxes
predicted by the analytical model using the parameters for the data sets 
{\it DATA\_SPECTRO\_CAL} (solid line) or {\it DATA\_INT\_CAL}
(dashed line), with the MIDI
flux (error bars), the ISO/SWS spectrum (blue line), and the
photometric data (open triangles).}
\label{SED_pi_ana}
\end{figure}

The model provides in both cases good quality fits ($\chi^{2}/(N-p)=\chi^{2}_{\rm red}$$<$0.6, where $N$ is the number of measurements and $p$ is the number of adjustable parameters); however, the parameter values are relatively
different for the two data sets, especially for the central star
diameter. Because this parameter is the most constraining for the
stellar spectrum, one can easily distinguish which of the parameter
sets provides the most reliable fit to the spectrum of the star 
by superimposing the analytical flux model generated with the 2
parameter sets onto the spectro-photometric data of the star.
Figure~\ref{SED_pi_ana} compares the \object{$\pi^{1}$\,Gruis} fluxes predicted by
the analytical model using the parameters for the data sets {\it DATA\_SPECTRO\_CAL} 
or {\it DATA\_INT\_CAL}, with the MIDI
flux, the ISO/SWS spectrum, and the
photometric data. It is very clear from this figure
that the parameter set providing the best fit to the spectrum is the
{\it DATA\_SPECTRO\_CAL} one (see Table~\ref{pi_gru_ana_para}). \\

The stellar diameter obtained from the fit of the analytical model is $\phi$=21.6~mas. This corresponds to R$_{\rm \star}$=357~R$_{\odot}$ for the adopted Hipparcos distance of D=153~pc. In a study of the angular size of a sample of S and C stars, \citet{vanb97} found that, for non-Mira stars, S-type stars are on average 110~R$_{\odot}$ larger than M-type stars. Since the relationship between radius and $V-K$ color derived by \citet{vanb99} applies to M stars, the corresponding radius of \object{$\pi^{1}$\,Gruis} is set to 359~R$_{\odot}$, a value very close to the one obtained from the fit of the analytical model.
In Table~\ref{pi_gru_ana_para}, the luminosity of the central star has been added. This one is deduced from 
the effective temperature of the central star (T$_{\rm eff}$=3000~K), its angular diameter
($\phi_{\rm \star}$=21.6$\pm$0.3 mas), and its distance (D=153$\pm$20 pc),

\begin{equation}
L_{\rm \star} = 4 \pi \left(\frac{\phi_{\rm \star}}{2} D \right)^{2} \sigma T_{\rm eff}^{4} \approx (9300\pm2700) L_{\odot},
\end{equation}
where $\sigma$ is the Stefan-Boltzmann constant. The corresponding
apparent bolometric magnitude of \object{$\pi ^{1}$\,Gruis} is 0.8$\pm$0.6, in
agreement with the value of 1.07 found by \citet{vane00b}.\\ 
Another significant parameter value is the density power-law coefficient. With an $r^{-1.7}$ density distribution, this suggests a decrease in mass-loss rate with time, in line with the finding of \citet{chiu06} that the mass-loss rate of \object{$\pi^{1}$\,Gruis} has probably decreased over the past 90 years. This set of parameters will thus be used
as input for the DUSTY model described in Sect.~\ref{dusty_code}.

\begin{table}[tp]
\caption{Parameters of \object{$\pi ^{1}$\,Gruis} deduced from the fit of the
analytical visibility model to the data sets \textit{DATA\_SPECTRO\_CAL}.}
\label{pi_gru_ana_para}
\begin{minipage}[h]{10cm}
\begin{tabular}
[c]{cc} \hline\hline
Parameter  & Value \\ \hline
Effective temperature (K) : T$_{\rm eff}$ &  3000 \\
Luminosity (L$_{\odot}$) : L$_{\rm \star}$ & 9300$\pm$2700 \\ 
Central star diameter (mas) : 2 $\varepsilon_{\rm \star}$ &  21.6$\pm$0.3 \\
Shell inner radius (mas) : $\varepsilon_{\rm in}$ &  159$\pm$10 \\
Inner boundary temperature (K) : T$_{\rm in}$ &  776$\pm$25 \\
Dust bump wavelength ($\mu$m) : $\lambda_{\rm c}$ &  11$^{*}$ \\
Bump amplitude ($\times$10$^{-4}$) : $A$ &  8.3$\pm$0.6 \\
Bump width ($\mu$m) : $\Delta\lambda$ & 2.8$\pm$0.3 \\
Bump optical depth ($\times$10$^{-3}$) : $\tau_{\rm 11 \mu m}$ &   1.5$\pm$0.2 \\ 
Density power law coefficient : $p$ &1.7 \\ 
Spectral index : $\beta$ &  1 \\
$\chi^{2}_{\rm red}$ &  0.55 \\
\hline
\end{tabular}
\end{minipage}
$*$ : fixed at pre-determined value.\\
\end{table}

\subsection{Numerical modeling}
\label{numerical_modeling}

\subsubsection{The DUSTY code}
\label{dusty_code}

The DUSTY code \citep{ivez99} was used to generate synthetic
spectrally-dispersed (8-13 $\mu$m) flux and visibilities to confirm
and complete the parameter set found with the analytical model (see Table~\ref{pi_gru_ana_para}). DUSTY
is a public-domain simulation code that solves the problem of radiation
transport in a circumstellar dusty environment by integrating the
radiative transfer equation in plane-parallel or spherical geometries
\citep{ivez97,ivez99}.

The DUSTY model integrates the MARCS synthetic stellar spectrum of the
central star (see Sect.~\ref{atmos_pigru}) and considers as
input parameters a part of those deduced from the fit of the
analytical model: T$_{\rm eff}$, L$_{\rm \star}$, T$_{\rm in}$, $\tau_{\rm 11 \mu m}$, and $p$ (see Table~\ref{pi_gru_ana_para}). DUSTY also requires the grain size distribution as input so we chose a standard MRN grain size
distribution as described by \citet{math77}: $n(a)$ $\propto$
$a^{-3.5}$ with minimum and maximum grain sizes of 0.005 $\mu$m and
0.25 $\mu$m, respectively.

Therefore, the analytical model allows one to reduce the number of
DUSTY input parameters from 8 to 2 (namely the chemical composition of
the dust and the thickness of the shell). Indeed, the chemical
composition of the dust is related to the value of the spectral index
$\beta$. However, this value encapsulates many characteristics related
not only to the chemistry of the grains but also to their nature
(amorphous or crystalline), their shapes, and their sizes.\\
The thickness of the shell was found to be the least accurate output
parameter, with a lower limit of 60 stellar radii, corresponding to
the beam size of an AT. In fact, varying this value from 60 to 1000
times the stellar radius does not have a significant impact on the
model. This is related to the low mass-loss rate of \object{$\pi^{1}$\,Gruis}
(4.6$\times$10$^{-7}$ M$_{\odot}$\,yr$^{-1}$) leading to a very low
dust density in the outer layers. Since the N band is not well-suited
to determining the outer radius of the dust shell, this one could 
extend to the inner radius of the flared disk located at 125 stellar radii
\citep{chiu06}. \\

\subsubsection{Dust chemical composition}
\label{chem_comp}

The last free parameter to determine for the DUSTY
modeling is the chemical composition of the dust shell. However, this
parameter can be constrained from the shape of the mid-infrared
spectrum.

The classification by \citet{litl88,litl90} (hereafter referred to as
the LML classification) gives a sequence from the classic, narrow-emission 
feature at 10~$\mu$m (classified as silicate) to features
with progressively stronger contributions at 11~$\mu$m (silicate+ and 
silicate++, depending on its strength) to broad, low-contrast emission
peaking longward of 11~$\mu$m (Broad). In this classification,
\citet{litl88} investigate the LRS (IRAS Low-Resolution Spectrometer)
spectra of a large number of MS and S stars and suggest that the
unusual dust chemistry (caused by a C/O ratio near unity) might
produce unusual dust excess. \citet{litl88} create a separate class
of silicate emission, the S class, which they associate with S stars
and describe as having a peak emission around 10.8~$\mu$m, which is
the case for \object{$\pi^{1}$\,Gruis} (see
Fig.~\ref{pigru_spectra_Jy}). However, later analysis by
\citet{chen93} and \citet{sloa95} concluded that the LRS spectra of S
stars are not significantly different from those of M
giants. \citet{sloa98} noticed that MS and S stars, having a C/O ratio
near unity, produce circumstellar shells dominated by alumina
grains. They suggest that, when the C/O ratio approaches unity, the
formation of CO leaves little oxygen available for grain formation. As
gaseous material moves away from the central star and cools, alumina
grains will form before silicate grains because of the higher
condensation temperature of alumina. \citet{sloa98} also suggest that
stars with C/O ratios near unity may exhaust the supply of available
oxygen with the formation of CO gas and alumina grains, preventing
silicates from forming. 

\begin{figure*}[tbp]
\begin{center}
\includegraphics[height=5cm]{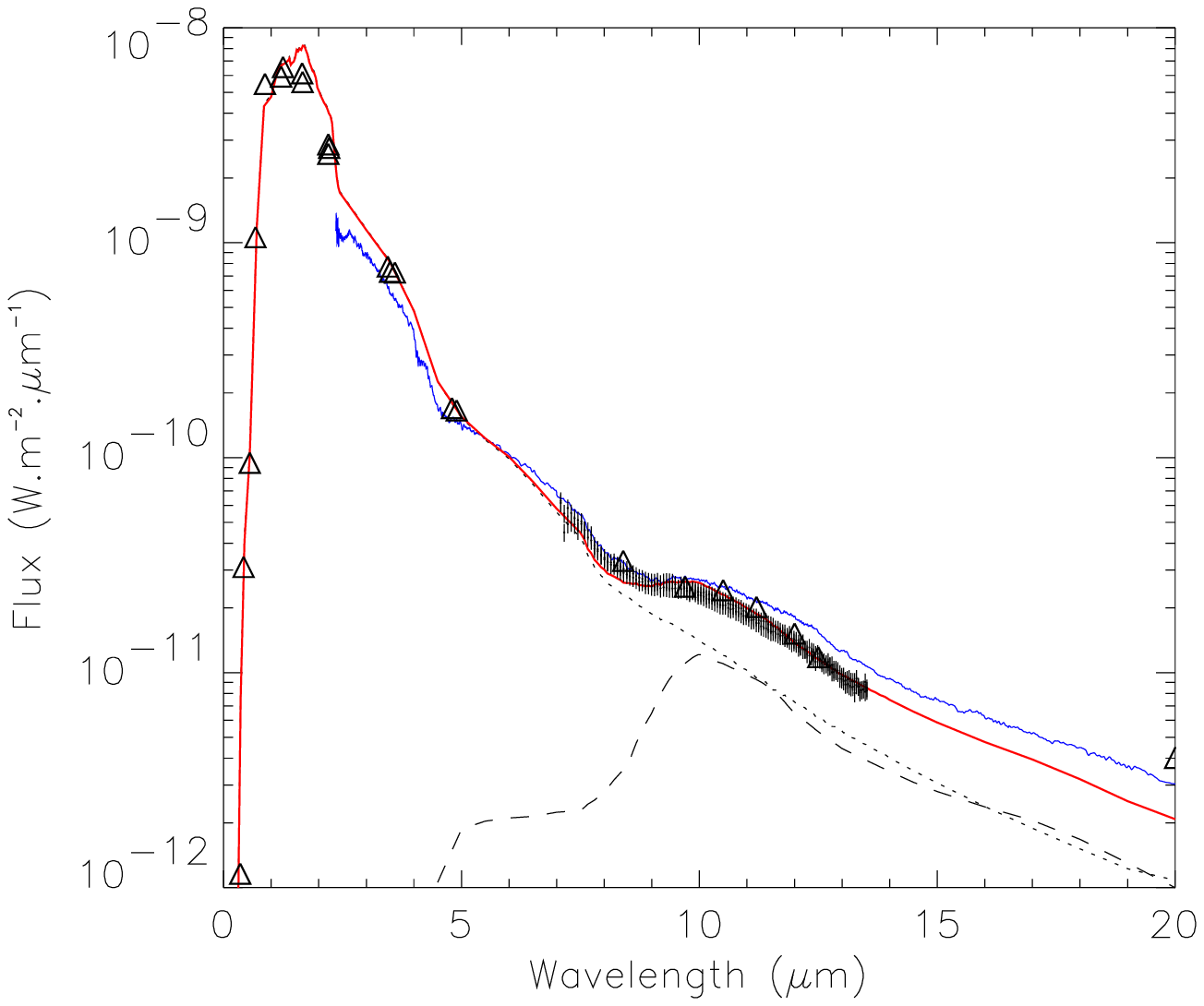} 
\hspace{0.5cm}
\includegraphics[height=5cm]{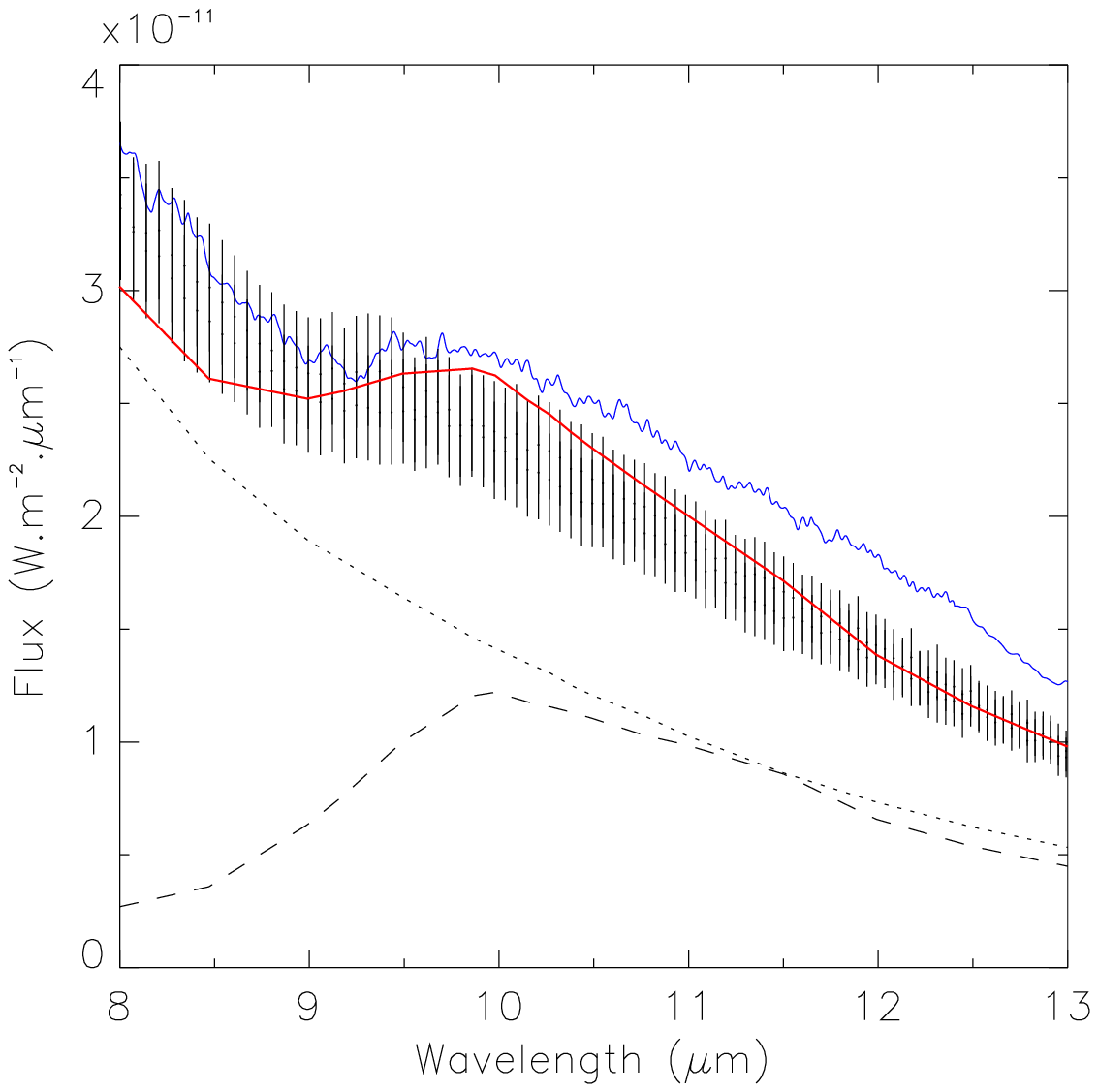} 
\begin{minipage}[c]{30cm}
\vspace{0.5cm}
\includegraphics[height=5.6cm]{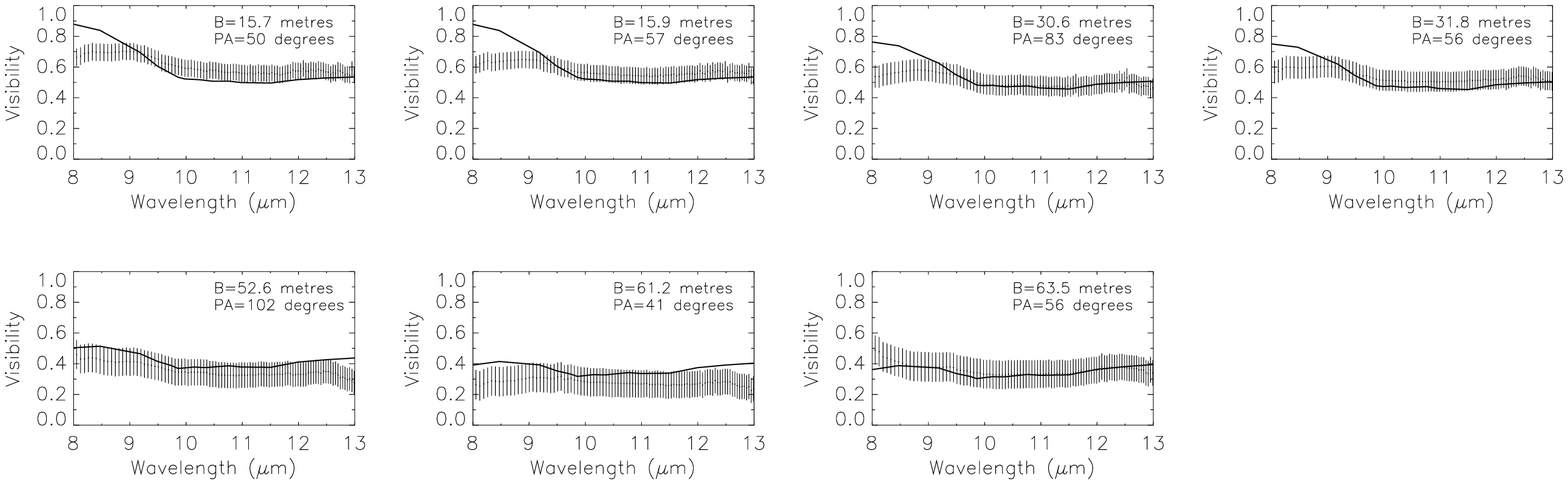}
\end{minipage}
\end{center}
\caption{Top-left: best fit of the dust shell model spectrum (red
  line) on the MIDI flux (error bars), the ISO/SWS spectrometric data
  (blue line), and the photometric data (open triangles) of
  \object{$\pi^{1}$\,Gruis}. The central star contribution (dotted line) and the
  dust shell contribution (dashed line) are added. 
Top-right: close-up view of the best fit of the dust feature. Bottom:
the corresponding model visibility (solid line) superimposed on the MIDI
visibilities (error bars) for the seven projected baselines.}
\label{pi_gru_dusty_1shell}
\end{figure*}

Figure~\ref{pigru_spectra_Jy} shows the broad dust emission feature from
9 to 13 $\mu$m peaking at 10.5 $\mu$m, which originates from the
optically and geometrically thin shell of \object{$\pi^{1}$\,Gruis}. This
spectral shape is well-known and corresponds to one phase of
the \textit{silicate dust sequence} scenario (hereafter referred to as
the SDS scenario) introduced by \citet{sten90}. In this scenario,
grains form first from condensing alumina material. Then, as the
grains evolve, silicate material begins to dominate the
emission. Fresh silicate material would be crystalline and exhibit
rather narrow emission components at 10 and 11 $\mu$m, while grains in
more extended shells would be amorphous and exhibit the classic
silicate feature. The S star \object{$\pi^{1}$\,Gruis} belongs to the bottom end of the SDS, 
classified as SE2 by \citet{sloa98}. This lower silicate
dust sequence is populated by spectra with broad, low-contrast
emission features, originating from optically thin shells of alumina
dust.\\
In this context, we first tried to reproduce the mid-infrared spectral
shape emission with a pure alumina shell using porous amorphous 
Al$_{2}$O$_{3}$ of \citet{bege97}\footnote{Note that the data
of \citet{bege97} only cover emission between 7.8 and
500~$\mu$m. DUSTY assumes a constant refractive index in the
unspecified shorter wavelength range, with a value equal to the
corresponding first point of the tabulation.} retrieved from the
Database of Optical Constants of the Laboratory Astrophysics Group of
the AIU
Jena\footnote{\tt{http://www.astro.uni-jena.de/Laboratory/Database/}}. This
composition failed to fit the spectrometric data. Spectra of SE2 class
stars do not exhibit any emission peak at 10~$\mu$m. However, the
ISO/SWS and MIDI spectra of \object{$\pi^{1}$\,Gruis} show that the emission feature
begins to peak at 10~$\mu$m, ending beyond 12~$\mu$m (see
Fig.~\ref{pigru_spectra_Jy}). Therefore, in spite of the SDS 
classification of \object{$\pi^{1}$\,Gruis} as SE2, it seems that this
star is close to the \textit{structured} silicate emission class
(SE3-SE6). This class is intermediate between the \textit{silicate}
and the \textit{broad} class stars implying that the model contains a
mixture of silicate and alumina grains, as discussed by
\citet{lore00}. Consequently, we decided to work with mixtures of
aluminum oxide and warm silicate \citep{osse92}, ranging from 100$\%$
alumina grains to 100$\%$ silicate grains in 10$\%$ increments. We
find a good fit (see Fig.~\ref{pi_gru_dusty_1shell}) with a ratio of
30$\%$ alumina and 70$\%$ silicate grains (see Table~\ref{tab_pi_gru_dusty_1shell}).

\begin{table}[tbp]
\caption{Parameters for \object{$\pi ^{1}$\,Gruis} deduced from the fit of the
DUSTY model to the interferometric and spectro-photometric
data.}
\label{tab_pi_gru_dusty_1shell}
\begin{minipage}[h]{10cm}
\begin{tabular}
[c]{cc} \hline\hline
Parameter & Value \\ \hline
Effective temperature (K)  & 3000$^{+}$ \\ 
Luminosity (L$_{\odot}$) & 8700$^{+}$ \\
Central star diameter (mas) & 21.0 \\
Shell inner radius (mas)  & 149.5 \\
Inner boundary temperature (K)  & 700$^{+}$ \\
Grain chemical composition  & 70$\%$ WS + 30$\%$ Al. \footnotemark[1]\\
Density power law coefficient  & 1.7$^{+}$ \\
Grain size distribution  & MRN$^{*}$ \\
Geometrical thickness ($\varepsilon_{\rm \star}$)& 60\footnotemark[2] \\
11 $\mu$m optical depth ($\times$10$^{-3}$) & 8.3$^{+}$ \\
\hline
\end{tabular}
\end{minipage}
$*$ : fixed at pre-determined value.\\
$+$ : derived from the analytical model (see Table~\ref{pi_gru_ana_para}).\\
\footnotemark[1]{WS stands for warm silicates and Al. for alumina.}\\
\footnotemark[2]{This value is not well defined (see text).}
\end{table}

The 11 $\mu$m global optical depth found with
the analytical method ($\tau_{\rm 11\mu m}$=(1.5$\pm$0.2$)\times$10$^{-3}$; Table~\ref{pi_gru_ana_para})
appears to be somewhat too small for DUSTY, which requires about 5 times more optical
depth to yield a good fit of the IR
spectrum. This may be due to the analytical model
assuming that the 11~$\mu$m feature may be approximated by a Gaussian
function, whereas Fig.~\ref{pigru_spectra_Jy} shows that this excess
is clearly asymmetric.

The temperature at the inner boundary of the shell (700~K;
Table~\ref{tab_pi_gru_dusty_1shell}) is below the aluminum oxide
(T$_{\rm cond}^{\rm Al_{2}O_{3}}$$\approx$1500-1800~K) and silicate
(T$_{\rm cond}^{\rm Si}$$\approx$1000-1500~K) condensation temperatures. Such
low temperatures in the shells of evolved stars were first suggested
by \citet{rowa82} and later by \citet{onak89} and \citet{simp91},
based on studies of the IRAS/LRS data. \citet{onak89} argue that these
low temperatures cannot be understood in the framework of homogeneous
nucleation theories and propose, as a possible explanation, that
mantle growth on pre-existing aluminum oxide particles is the major
process of silicate formation. 
Extending the temperature at the inner boundary of the shell 
to alumina condensation temperature ($\sim$1500 K) can also explain 
both the spectrum and visibility of the star. However, assuming a mass-loss rate value of 4.6$\times$10$^{-7}$M$_{\odot}$\,yr$^{-1}$ for \object{$\pi^{1}$\,Gruis} \citep{jori98} and a dust-particle specific density of 3 g cm$^{-3}$, 
the value of the gas-to-dust ratio of the model is 60. We obtain a value of 165 for the inner boundary temperature of 700 K, which agrees with the gas-to-dust ratio of 158$\pm$13 obtained by \citet{knap85} for oxygen stars.\\

Figure~\ref{pi_gru_dusty_1shell} shows the best interferometric and
spectro-photometric fit of the DUSTY model corresponding to the
parameter set of Table~\ref{tab_pi_gru_dusty_1shell}.
The model appears to match the observed spectra well and the MIDI
visibilities fairly well. However, while the spectral fit of the DUSTY
model is good overall but especially so in the mid-infrared band (8-13
$\mu$m), the model visibilities are over-evaluated for the two
$\sim$15 and 30 m projected baselines in the 8-9 $\mu$m spectral
band (see Fig.~\ref{pi_gru_dusty_1shell}). This reveals that the model is not extended enough for these
baselines at these wavelengths. This can be explained by the very low 
optical depth in that spectral range ($\bar{\tau}$(8-9~$\mu$m)=2.5$\times$10$^{-3}$) leading to a
domination of the central star flux over the flux of the dust shell. In the 8-9 $\mu$m spectral
range, the model therefore has values very close to the central star's 
uniform-disk intensity distribution, leading to an increase in the
visibility. Moreover, one can see that the 8-9 $\mu$m 
visibilities decrease with the increase in the baseline lengths, which is
understandable in the context of an equivalent uniform-disk visibility
function. \\
A way to increase the spatial extension of the model between 8 and 9 $\mu$m is
to add a molecular shell between the photosphere of the star and
the inner boundary of the dust shell. In the next section, we
investigate the possibility of adding such a molecular shell in the model.

\subsection{Addition of a warm H$_{2}$O+SiO molecular shell}
\label{double_shell}

Recent work reveals that diameters of the Mira and non-Mira stars appear
systematically larger than expected in the near-infrared
\citep{menn02,ohna04b,perrin04,weiner04} and in the mid-infrared
\citep{weiner00,weiner03a,weiner03b,weiner04,ohna04b,ohna05}. Such an
increase cannot be attributed only to dust shell
features, but also to the possible existence of extended gaseous shells 
close to the stellar photospheres.\\
The analyses of the spectra obtained with
ISO/SWS have revealed that the spectra of Mira variables are well-reproduced 
by water vapor shells extending to 2-4 stellar radii
with gas temperatures of 1000-2000 K and column densities of
10$^{20}$-10$^{22}$ cm$^{-2}$ \citep{tsuj97,yama99,mats02}. Since \object{$\pi^{1}$\,Gruis} 
is a non-Mira star, we should expect different parameter
values than those for Mira variables.\\
The single dust shell modeling with
DUSTY (see Fig.~\ref{pi_gru_dusty_1shell}) suggests that we need to
have a contribution from a molecular shell only in the 8-9~$\mu$m
spectral band. Such a molecular shell, also called "MOLsphere"
\citep{tsuj00}, is favored by the low surface gravity (log $g$=0; see
Table~\ref{stellar-parameters-pi}) and high luminosity ($\sim$9000
$L_{\odot}$) of the star, allowing the levitation of the upper
layers. The temperatures in this levitated matter can be low enough
for molecules to form \citep{verh06}.

Because this molecular shell has a larger extent than the
central star, the former is resolved with smaller baselines so as to
decrease the visibility amplitudes in that spectral range. At the same
time, the molecular shell must make a negligible contribution to the
total flux in the 9 to 13 $\mu$m spectral range relative to the
dust shell, to recover the good fit that we found with the
single dust shell model at these wavelengths (see
Fig.~\ref{pi_gru_dusty_1shell}).\\

Water vapor (H$_{2}$O) and silicon monoxide (SiO) are
the main molecules contributing to absorption and causing the apparent
mid-infrared diameter to be larger than the continuum diameter 
\citep{ohna05,verh06,perrin07}.   
The H$_{2}$O lines are present across the entire N band range, while
most SiO lines are located between 8 and 10~$\mu$m.

The model of molecular shell that we chose is the same as the one
used by \citet{ohna05} and consists of H$_{2}$O and SiO gas with
constant temperature and density, extending from the central star
radius up to the external boundary $\varepsilon_{\rm mol}$. The input
parameters of the model are the temperature of the molecular shell 
(T$_{\rm mol}$), the column densities of H$_{2}$O and SiO 
(N$^{\rm H_{2}O}_{\rm col}$ and N$^{\rm SiO}_{\rm col}$, respectively) 
in the radial direction, and the geometrical extension of the shell 
($\varepsilon_{\rm mol}$).\\ 
In this model, we first calculate the line opacity due to
H$_{2}$O and SiO with a wavenumber interval of 0.04 cm$^{-1}$ over a
wavelength range between 8 and 13 $\mu$m. We adopt a Gaussian line profile 
with an FWHM of 5~km\,s$^{-1}$, which
represents the addition of a gas thermal velocity of a few
km\,s$^{-1}$ (V$_{\rm T}$=1.5(T$_{\rm mol}$/300 K)$^{0.5}$ km\,s$^{-1}$) and the
micro-turbulent velocity in the atmosphere of \object{$\pi ^{1}$\,Gruis}
(3~km\,s$^{-1}$; see Table~\ref{stellar-parameters-pi}) and assumes that
the molecular gas is in local thermodynamical
equilibrium. 
The list of H$_{2}$O lines was taken from the HITEMP
database \citep{roth97} while
that of the fundamental bands of SiO was generated from the
Dunham coefficients given by \citet{lova81} and the dipole moment
matrix elements derived by \citet{tipp81}.\\

The resulting intensity distribution of the object is composed of the
central star with a MARCS model atmosphere, the molecular 
shell, and the dust shell. The analytical expression of the model is
given by the expressions below:

\begin{eqnarray}
\lefteqn{ I^{\rm \star}_{\lambda}(\varepsilon)=I^{\rm marcs}_{\lambda}e^{-\tau_{\rm mol}}e^{-\tau_{\rm dust}} \ \Pi\left(\frac{\varepsilon}{2\varepsilon_{\rm \star}}\right) } \\
\lefteqn{ I^{\rm mol}_{\lambda}(\varepsilon)=B_{\lambda}(T_{\rm mol}) \ e^{-\tau_{\rm dust}}\left[1-e^{-\tau_{\rm mol}}\right] \ \Pi\left(\frac{\varepsilon}{2\varepsilon_{\rm mol}}\right) } \\
\lefteqn{ I^{\rm dust}_{\lambda}(\varepsilon)=B_{\lambda}(T_{\rm dust})\left[1-e^{-\tau_{\rm dust}}\right] \ \Pi\left(\frac{\varepsilon}{2\varepsilon_{\rm out}}\right) }
\end{eqnarray}
where $\tau_{\rm mol}$ and $\tau_{\rm dust}$ are the optical depths of the
molecular and dust shells, respectively, and $\Pi$ is the
uniform disk distribution ($\Pi(x)$ = 1 if $-1/2<x<1/2$, zero
otherwise).\\ The corresponding flux and visibility, obtained by the
Hankel transform of the intensity profile, are then spectrally
convolved to match the low spectral resolution (30) of the
MIDI prism. Using the above model and without changing the
parameter values found for the thin dust shell (see
Table~\ref{tab_pi_gru_dusty_1shell}), we search for the best
combination of input parameters (T$_{\rm mol}$, $\varepsilon_{\rm mol}$, 
N$^{\rm H_{2}O}_{\rm col}$, and N$^{\rm SiO}_{\rm col}$) to improve the spectro-interferometric fitting in
the 8-9 $\mu$m spectral range. The best-fitting values are presented
in Table~\ref{Tmol_rmol_Ncol}.

\begin{table}[tbp]
\centering
\caption{Best combination of input values for the determination of the molecular shell.}
\label{Tmol_rmol_Ncol}
\begin{tabular}
[c]{cccc} \hline\hline
\vspace{0.1cm}
Component & T$_{\rm mol}$ (K) & $\varepsilon_{\rm mol}$ ($\varepsilon_{\rm \star}$) & N$_{\rm col}$ (cm$^{-2}$) \\
H$_{2}$O & 1000 & 4.4 & 5$\times$10$^{18}$ \\
SiO & - & - & 2$\times$10$^{18}$ \\
\hline
\end{tabular}
\end{table}

These values correspond to an extended, optically thin ($\tau_{\rm mol}$$<$0.1) molecular shell.
Such a diluted molecular shell can be connected to the
nature of \object{$\pi ^{1}$\,Gruis}. First, because the C/O ratio is close to
unity for S stars (C/O=0.97 for \object{$\pi^{1}$\,Gruis}; see
Fig.~\ref{Color-Color}), the last oxygen-bearing molecules disappear
\citep{ferr02}. Second, due to its classification as a non-Mira semi-regular
variable, for which the variability amplitudes are much smaller than
for Mira variables, a less dense molecular shell is indeed
expected. From these arguments, it seems reasonable that this star
exhibits weak spectral features from the molecular shell.

\begin{figure*}[tbp]
\begin{center}
\includegraphics[height=5.9cm]{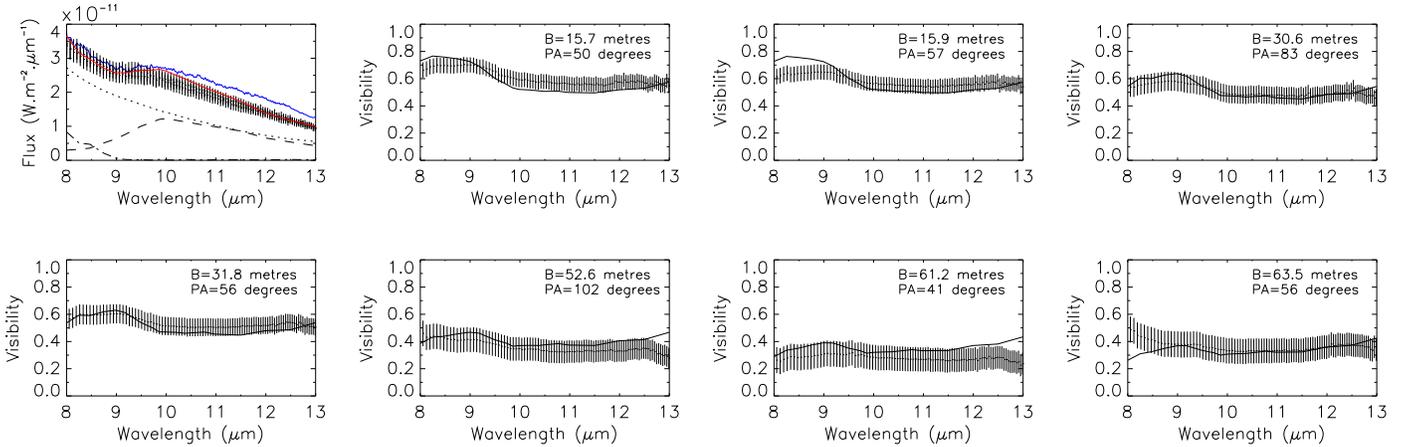}
\end{center}
\caption{Top-left: best fit of the proposed H$_{2}$O+SiO molecular 
shell and silicate+alumina dust shell model spectrum (red line) on the MIDI flux
(error bars) and the ISO/SWS spectrometric data (blue line) of
\object{$\pi^{1}$\,Gruis}. The central star contribution (dotted line), the
molecular shell contribution (dot dashed line), and the dust shell
contribution (dashed line) are added. 
The other figures show the corresponding model visibility (solid line) superimposed
on the MIDI visibilities (error bars) for the seven projected
baselines.}
\label{pi_gru_dusty_mol_dust}
\end{figure*}

Figure~\ref{pi_gru_dusty_mol_dust} shows our best spectrometric and
interferometric fit to the optically thin H$_{2}$O+SiO molecular shell and 
optically thin silicate+alumina dust shell model using the parameter sets of
Tables~\ref{tab_pi_gru_dusty_1shell} and \ref{Tmol_rmol_Ncol}.

This figure reveals that the spectrometric, as well as the visibility
measurements are now well-reproduced over all the 8-13 $\mu$m spectral
band for all projected baselines. In particular, the model
visibilities have decreased in the 8-9~$\mu$m spectral band compared
to the single dust shell model (see
Fig.~\ref{pi_gru_dusty_1shell}). This is due to the correlated flux
contribution of the molecular shell leading to a more extended
object at these wavelengths. With a size of 4.4 stellar radii, the molecular shell 
is almost completely resolved with baselines larger than 30 meters. 
For this reason, its 
contribution to the correlated flux is greater for the 15- and 30-meter 
projected baselines than for the 60-meter one. The optical depth of the molecular shell outside the 8 to 9~$\mu$m range is
ten times less than in this band. Therefore, the correlated flux contribution of
the molecular shell is almost negligible at these wavelengths,
preserving the good fit from the single dust shell model.\\

From this last model, we can now infer the spatial circumstellar
structure of \object{$\pi^{1}$\,Gruis}. It consists of an optically thin warm
H$_{2}$O+SiO molecular shell extending from the stellar photosphere
out to 4.4 stellar radii, then an optically thin alumina-silicate dust
shell beginning at about 14 stellar radii, and finally a slowly
expanding flared disk with a central cavity of radius about 125
stellar radii, determined by \citet{chiu06} in the millimetric spectral
range.

\section{Asymmetries of the circumstellar material and binarity }
\label{asymm}

The bipolar outflow perpendicular to the flared disk was discovered by
\citet{saha92} and subsequently studied by \citet{knap99a} and
\citet{chiu06}. A similar structure combining disk and jet was 
observed by \citet{knap97} in another very
evolved AGB star, \object{V\,Hya}, whose binary nature was subsequently
discovered by \citet{knap99b}.  As argued by \citet{hugg07}, the
binarity seems to be a necessary condition creating a disk/jet
structure in evolved mass-losing stars. Although \object{$\pi^{1}$\,Gruis}
has a G0V companion located 2.71$\arcsec$ away, this companion falls
within the expanding disk \citep{chiu06}, a strange situation that 
has not been encountered in, for example, binary post-AGB stars, where the
companion falls well within the cavity interior to the disk \citep{wael96, domi03}. 
Therefore, \citet{chiu06} wonder whether
there `might be a much closer companion that has so far escaped
detection and currently is the source of the high-velocity outflow'.\\
\citet{maka05} and \citet{fran07} find possible evidence for such a
close companion from the discrepancy between the Hipparcos and Tycho-2
proper motions. The Hipparcos proper motion relies on measurements only 
spanning 3~years, whereas the Tycho-2 one uses positional data
spanning more than a century. Any discrepancy between these
short-term and long-term proper motions is thus indicative of an 
unrecognized orbital motion in addition to the 
true proper motion. The discrepancy is very significant in the case of
\object{$\pi^{1}$\,Gruis}. According to the analysis of \citet{fran07}, these
so-called `$\Delta \mu$ binaries' correspond to systems with orbital
periods in the range 1500-$10^{4}$~days, much shorter than the
6000~years period inferred by \citet{knap99a} for the G0V companion
2.7$\arcsec$ away.\\

We now estimate the likely range for the orbital separation of the
putative close companion and compare it to the size of the various
shells and to the stellar radius. The period range for $\Delta\mu$
binaries with known orbital elements are from 1500 to 10000~days,
according to Fig.~4 of \citet{fran07}. Adopting a total mass of
2~M$_{\odot}$ for the system, the corresponding range in relative
semi-major axes are 3.2 to 11.4~AU, or 20.8 to 74.4~mas (for the
parallax of 6.5~mas). Adopting 3~M$_{\odot}$ instead yields 3.7 to
13.1~AU or 24 to 85~mas.  These values must be compared to the
\begin{itemize}
\item stellar radius = $\varepsilon_{\rm star}$ = 10.5 mas
\item H$_{2}$O+SiO molecular radius = 4.4 $\varepsilon_{\rm star}$ = 46.2 mas
\item inner dust shell radius = 14.2 $\varepsilon_{\rm star}$ = 149.5 mas.
\end{itemize}
It is seen that only the longest periods and highest total masses are
compatible with the shell sizes. One can wonder what in this case would be the impact of a 
companion orbiting at about 80~mas from the central star on the molecular and dust shell. 
As the extended dust shell is over-resolved, only the molecular shell could provide some hints of the presence of a perturbation. However, the mass-loss rate is so low that a high dynamic range is required to detect the asymmetries from the faint contribution from the molecular and dusty material. Such a dynamic range cannot correctly be achieved with the current MIDI capabilities. Nevertheless, the constraints provided by the differential phases that do not exhibit any obvious departure from spherical symmetry (see Fig.~\ref{phase_pigru}) have to be considered, since it is more sensitive to any departure from centro-symmetry.\\ 
The $\Delta\mu$ method offers
no way to infer the position angle of the companion. However, if it is
powering the jet, its orbital plane is likely to be coincident with
that of the flared disk. The MIDI interferometric $uv$ coverage (PA=41 to 102$^{\circ}$) is oriented in the direction of the
flared disk (and thus along the orbital plane), as deduced from Fig.~4 in \citet{chiu06}. It is then difficult
to envisage the presence of a close companion without any detection of asymmetries. 
This could be the case if the dust shell were not heated by
the companion, which must therefore be on the lower main sequence (cool
and faint). \\

Beyond there being no detection of asymmetries as
shown by the zero-differential phase (see Fig.~\ref{phase_pigru}), there 
is a more fundamental information inferred from the interferometric observations in 
Fig.~\ref{symmetry_pigru}. There is no obvious indication that the star alone could drive the jet and shape 
its surrounding. Notwithstanding this fundamental constrain, we are left with two possibilities, each of them invoking the influence of a companion:\\

(i) There is indeed a close companion (cool and faint) to
power the jet, but then why are there no departures from
spherical geometry close to the star, related to the presence of an accretion disk?\\

(ii) The close $\Delta\mu$ companion is spurious, but then the jet
must be powered by the far-away companion, and why is this
outer companion orbiting the disk without creating a gap in it, as would be evident in the millimetric observations? \\

Regarding item (i), one explanation could come from the close, 
optically thin environment of \object{$\pi^{1}$\,Gruis} leading to similar SEDs
for accretion disks and spherical dust shells, as discussed
by \citet{knap93} and \citet{vand94}. If we add to this argument the fact that the $uv$
coverage is preferentially oriented in the direction of the disk having an
opening angle larger than the interferometric angular coverage
($\sim$60$^{\circ}$), it is then easy to confuse this structure with a spherical shell.\\

Regarding item (ii), it must be mentioned that the $\Delta\mu$ method
might face problems when dealing with highly evolved stars and might
then yield spurious results. To substantiate this, notice that 
Table 8 of \citet{fran07} lists several other candidate $\Delta\mu$ binaries among
long-period variables and suggests that these detections are similarly
spurious, although the reason thereof remains unidentified.\\

Still, even in the absence of a close companion, the \citet{hugg07}
analysis of the jet/torus properties in several AGB stars (among which
\object{$\pi^{1}$\,Gruis}) and proto-planetary nebulae requires them to be
binaries. The 2.7$\arcsec$ companion to \object{$\pi^{1}$\,Gruis} must then be
the one responsible for driving the jet/torus structure. In that case
however, the conclusion of the CO data analysis by \citet{chiu06}, 
which locates the companion {\em within} the flared disk, is quite
surprising (when compared to situations prevailing in post-AGB systems
for instance). Such a situation cannot be stable in the long term since the 
companion should carve a gap in the disk. If correct, it indicates
that the disk is quite young, as confirmed from its kinematics. Taking into account a mean value of 11~km\,s$^{-1}$ for the wind
expansion velocity found by \citet{chiu06} and a distance of 153
parsec \citep{perr97}, we can estimate the typical transit time of the
dust shell to be about 10 years, while the flared disk age is about 350
years \citep{hugg07} with a strong decrease in the mass-loss rate
about 90 years ago \citep{chiu06}.\\

\section{Summary and conclusions} 
\label{conclusion}

The goal of this work was to perform a multi-wavelength and multi-spatial
resolution study of the close environment of \object{$\pi^{1}$\,Gruis} 
with the N band spectral range of MIDI. Our new high spatial-resolution observations of the environment of
\object{$\pi^{1}$\,Gruis}, consisting of MIDI/VLTI spectra and visibilities,
combined with the existing millimeter data, led to a more 
comprehensive picture of its circumstellar environment on different 
spatial and spectral scales.\\

A preliminary study of MIDI data shows that the mas-scale observations
do not reveal any strong departure from the sphericity of the
circumstellar environment, whereas arcsec-scale millimetric
observations by \citet{knap99a} and \citet{chiu06} show strong
evidence of a slow and dense outflow (v$\sim$15~km\,s$^{-1}$) and of a
fast bipolar ejection (v$\geq$90~km\,s$^{-1}$).
Even though the constraints provided by these interferometric observations are limited in terms of $uv$ coverage and dynamics, this suggests that the companion located at 2.7$\arcsec$ (400 AU at 153~pc,
i.e. far outside the MIDI interferometric Field-Of-View of about 175
AU) may have some influence on the shape of the large-scale CO environment and may possibly
launch the fast outflow that has been detected (cf. the models of
\citealt{mast99} and the discussion in \citealt{chiu06}).\\ 

The more detailed study starts with determining a synthetic
atmosphere model for \object{$\pi^{1}$\,Gruis} to obtain approximate 
physical characteristics of the central star. The dusty circumstellar
environment was initially analyzed 
with the optically thin analytical radiative transfer
model developed by \citet{cruz06}. 
This model allowed us to determine some parameter values related to the central star and its dust shell, which we used 
to generate a more refined model with the numerical
radiative transfer code DUSTY. This code helped us to determine the
chemical composition of the close circumstellar environment, which was one of the main aims of
this study. We show, from the literature, that this star 
belongs to one phase of the SDS scenario \citep{sloa98}, in which grains first form from condensing 
alumina material. We confirmed this tendency using spectro-interferometric
fitting and derived a ratio of alumina to silicate dust of
0.4, revealing that \object{$\pi^{1}$\,Gruis} is situated in an intermediate
class named the \textit{structured} silicate emission class.\\
We then included an optically thin H$_{2}$O+SiO
molecular shell extending from the photosphere of the star up to
4.4 stellar radii with a typical temperature of 1000 K to account for
the fact that the single dust shell model is not extended enough
in the 8-9~$\mu$m spectral range. 
The formation mechanism of molecular shells in
non-Mira semi-regular variables like \object{$\pi^{1}$\,Gruis}, in which the
variability amplitudes are much less than in Mira stars, can lead
to less dense MOLspheres. Moreover, as S stars have a C/O ratio close
to unity, neither oxygen-bearing nor carbon-bearing molecules are
abundant in the atmosphere, with most of carbon and oxygen atoms
locked up in CO molecules.\\

Complex structures revealed by millimetric observations
suggest that \object{$\pi ^{1}$\,Gruis} may be at the end of its evolution on the
AGB \citep{knap99a}. This evolutionary stage is confirmed by the position
of the star on the HR diagram of S stars (see Fig.7 of \citealt{vane98}). 
The large and diluted optically thin close
environment around evolved stars could thus be part of the natural
transition from the AGB to post-AGB and planetary nebulae stages. If this
is true, what are the mechanisms responsible for such physical
processes? Answers could come from long-term spectro-interferometric
observations of the star and better Fourier $uv$ coverage, which could be
provided by the second generation VLTI instruments. This should be carried 
out in close synergy with ALMA array observations, which should provide images 
in the millimeter domain at similar angular resolution.

\begin{acknowledgements}
We would like to thank Brandon Tingley for the English revision 
of the text and S.~Flament for his help on the data
handling. Mr. Sacuto benefits from a PHD grant from the Conseil
R\'egional Provence - Alpes - C\^ote d'Azur (France) managed by
ADER-PACA.
\end{acknowledgements}

\input{8306ref}

\newpage

\appendix

\section{Angular diameter of the calibrator \object{$\beta$\,Gruis}}
\label{angular_diam_calib}

The calibrator star \object{$\beta$\,Gruis} is a cool giant of spectral type
M5III without any detected circumstellar matter \citep{sloa98}. With
an estimated angular diameter around 27 mas \citep{bedd94,hera02},
this star clearly cannot be considered as an unresolved object for 
the baselines used in this study.

To determine the theoretical visibility of \object{$\beta$\,Gruis}, one
can consider this star as having a uniform disk intensity. Such a 
hypothesis is valid in the mid-IR because of the vanishing
limb-darkening coefficient at these wavelengths. At 10~$\mu$m, the
ratio between the limb-darkened angular diameter and the uniform disk
angular diameter is very close to one \citep{davi00}. The visibility of the
uniform disk is

\begin{equation}
\label{uniform_disk}
V_{\rm UD}\left(\frac{B_{\rm p}}{\lambda}\right) = 2\left\vert\frac{J_{1}\left(\pi\Phi_{\rm cal}\frac{B_{\rm p}}{\lambda}\right)}{\pi\Phi_{\rm cal}\frac{B_{\rm p}}{\lambda}}\right\vert,
\end{equation}
where $\Phi_{\rm cal}$ is the angular diameter of the calibrator
\object{$\beta$\,Gruis}, $B_{\rm p}$ the projected baseline, $\lambda$ the
observing wavelength, and $J_{1}$ the first-order Bessel function of
the first kind. \\

Previous spectrometric \citep{hera02} and interferometric
\citep{bedd94} studies of this star allowed angular diameters of 28.1 mas 
and 27.0 mas, respectively, to be derived, although these
evaluations bear an uncertainty of $\pm$3 mas. This corresponds to an
error bar on the equivalent uniform disk visibilities reaching 20$\%$
at high spatial frequencies. Moreover, this error comes on top of the
transfer function error bar of 10\% (see Sect.~\ref{obs_cal-process}). Using the observed angular
diameters of \object{$\beta$\,Gruis} would thus carry an uncomfortably large
error bar of about 30\%. We therefore chose instead to predict the
diameter of \object{$\beta$\,Gruis} by other means to reduce this error. Two
methods are considered in the following, one with a 
dedicated SED and the other a visibility model of the star.

\subsection{Spectrometric estimation}
\label{spectro_estimation}

A fit of a spectral energy distribution (SED) model to the
spectrometric measurements of the calibrator allows the
angular diameter of the calibrator \object{$\beta$\,Gruis} to be determined:

\begin{equation}
\label{SED_estimation}
\pi \times \left(\frac{\Phi_{\rm cal}}{2}\right)^{2} I^{\rm cal}_{\lambda}=F^{\rm cal}_{\lambda},
\end{equation}
where $\Phi_{\rm cal}$ is the angular diameter of the calibrator,
$I^{\rm cal}_{\lambda}$ the intensity of the calibrator from a MARCS model, 
and $F^{\rm cal}_{\lambda}$ the ISO/SWS spectrometric data flux of the
calibrator \object{$\beta$\,Gruis}.

We used the MARCS code \citep{gust75, plez92, plez93}, which
generates stellar atmospheres of cool evolved stars, solving radiative
transfer in a hydrostatic LTE environment using spherical
geometry. To deduce the closest MARCS model for the
calibrator \object{$\beta$\,Gruis}, we must know the stellar atmospheric
parameters. These parameters have been derived by \citet{judg86} using
high-resolution spectra from the \textsl{International Ultraviolet
Explorer (IUE)} satellite:

\begin{itemize}

\item ~T$_{\rm eff}$=3400$\pm$100K
\item ~log~$g$=0.4
\item ~[Fe/H]=0.0.

\end{itemize}

The last stellar atmospheric parameter is the C/O ratio. The 
carbon abundance ($\log \epsilon(^{12}$C)=8.4) is scaled to the 
hydrogen ($\log \epsilon$(H)=12) and oxygen ($\log \epsilon(^{16}$O)=8.73) 
average abundances for the sample of normal M giants derived 
by \citet{smit85}. These lead to a C/O ratio of 0.47 for \object{$\beta$\,Gruis}.\\

The U/B/V/R/I/K/L/M/[11]/[22]/[60] photometric data are taken from
\citet{john66,fein66,mend69,enge92}. Dereddening is not necessary, 
since the galactic latitude of the star (-58$^\circ$)
leads to zero visual extinction \citep{chen98}. Errors on the angular
diameter of the star can be estimated from the phases of the 2.36-45.39
$\mu$m spectrometric data from the ISO/SWS (Infrared Space
Observatory, Short-Wavelength Spectrometer) \citep{sloa03}. The 
spectra of this star were obtained on 1996 Apr 9 and
1997 May 7. Assuming a period of 37 days for the SRb variable with a V
amplitude of 0.34 mag, and using the ephemeris of \citet{oter06}, we
find that the ISO observations are taken at phases 0.61 (i.e. shortly
after the photometric maximum) and 0.99, at the photometric
minimum. Thus, the MIDI phases of this star 
are approximately intermediate between the two ISO/SWS phases,
except for the 2006 June 19 observation corresponding to phase
0.01 (see Table~\ref{journal}).

\begin{figure}[h!]
\begin{center}
\includegraphics[width=8.cm]{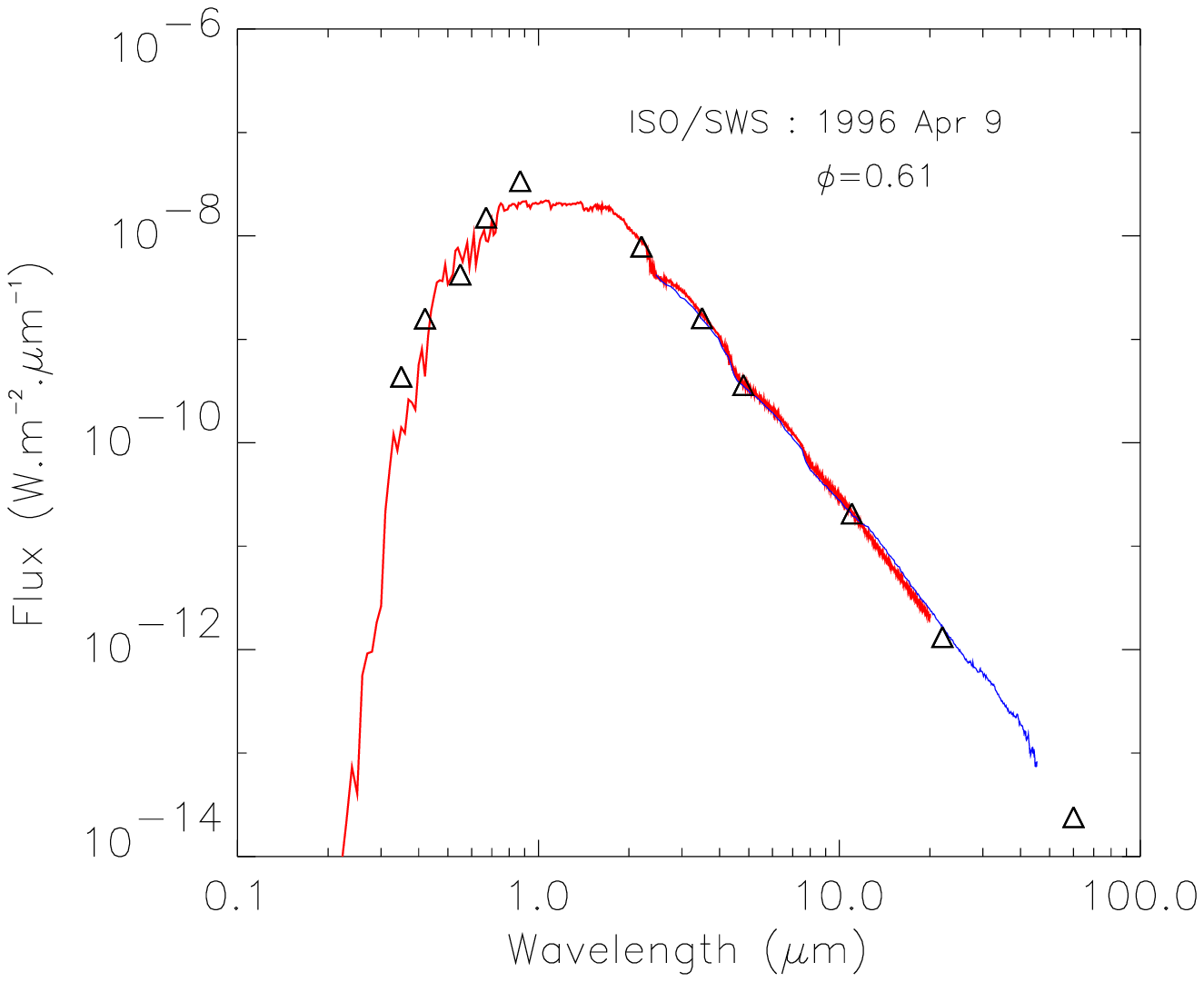}
\includegraphics[width=8.cm]{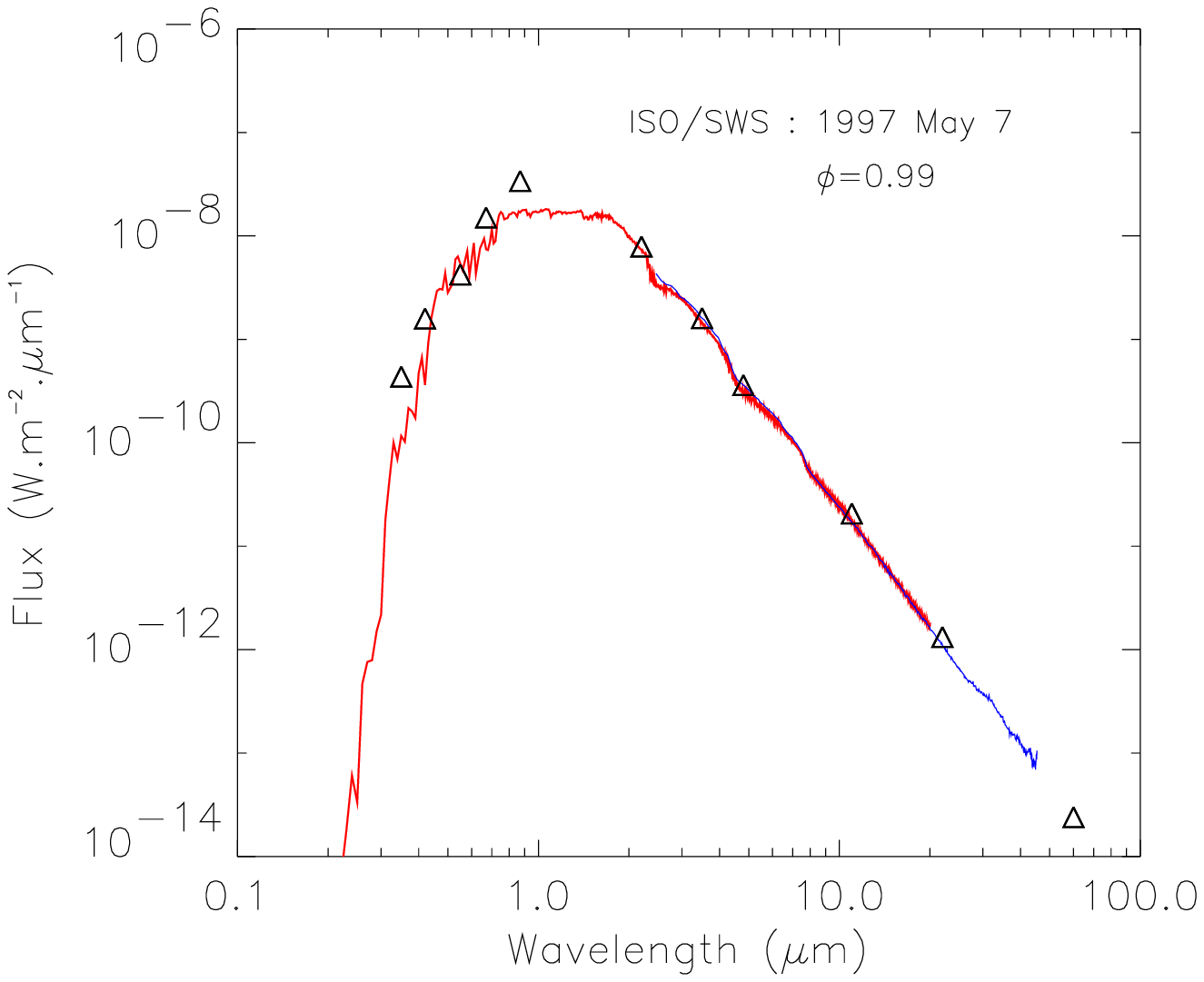}
\caption{Top: best fit of the MARCS model spectrum (red line) on the
calibrator \object{$\beta$\,Gruis} ISO/SWS data (blue line) obtained on 1996 Apr 9 (Phase
0.61). Bottom: same as the previous one for 1997 May 7 (Phase
0.99). The photometric (U/B/V/R/I/K/L/M/[11]/[22]/[60]) data of
\object{$\beta$\,Gruis} (open triangles) are superimposed on each figure.}
\label{SED_beta}
\end{center}
\end{figure}

Figure~\ref{SED_beta} presents the best fits of the MARCS model spectrum 
on each of the two ISO/SWS spectrometric observations (1996 Apr 9:
Phase 0.61 and 1997 May 7: Phase 0.99) yielding, from
Eq.~\ref{SED_estimation}, the angular diameter of the calibrator. Both
fits give an angular diameter of 28.0 mas for Phase 0.61 and 25.5 mas
for Phase 0.99, leading to a mean angular diameter of 26.8 mas with a
mean standard deviation of 1.3 mas. This error is about 2 times
smaller than the one found in the literature \citep{hera02} for
\object{$\beta$\,Gruis}, thus reducing the corresponding error bar on the
uniform disk visibility (from 0.6$\%$ at low spatial frequencies to
10$\%$ at high spatial frequencies).

\subsection{Interferometric estimation}
\label{interfero_estimation}

Another method for deriving the angular diameter of the calibrator
\object{$\beta$\,Gruis} consists of using observations of several
calibrators observed during the same nights but at a different sky
position. The journal of observations for these stars (see
Table~\ref{journal_2}) shows their spectral type and their angular
diameter with their errors calculated from the ESO VLTI Calibrators
Program \citep{rich05}, with the position, UT time, and the relative
projected baselines of the calibrated sources compared to the ones for
\object{$\beta$\,Gruis} of each observation.

\begin{table*}[tbp]
\caption{Journal of observations of the calibrators observed the same nights as \object{$\beta$\,Gruis}.}
\label{journal_2}
\vspace{0.3cm}
\begin{center}
\begin{tabular}{ccccccc}
\hline
\hline
{\tiny Star} & {\tiny Spectral type} & {\tiny Right ascension ; Declination} & {\tiny UT date \& Time} & {\tiny d$_{\rm UD}$[mas]} & {\tiny Base[m]} & {\tiny PA[deg]} \\
\hline
{\tiny \object{$\beta$\,Gru}} & {\tiny M5III} & {\tiny 22 42 40 ; -46 53 04} & {\tiny 2006-05-25 07:51:28} & {\tiny \ldots} & 60.2 & 32 \\
{\tiny \object{$\epsilon$\,Peg}} & {\tiny K2Ib} & {\tiny 21 44 11 ; +09 52 30} & {\tiny 2006-05-25 09:24:17} & 7.7$\pm$0.2 & 59.1 & 75 \\
{\tiny \object{$\beta$\,Gru}} & {\tiny M5III} & {\tiny 22 42 40 ; -46 53 04} & {\tiny 2006-05-25 07:51:28} & {\tiny \ldots} & 60.2 & 32 \\
{\tiny \object{$\eta$\,Sgr}} & {\tiny M3.5III} & {\tiny 18 17 38 ; -36 45 42} & {\tiny 2006-05-25 03:25:52} & 11.9$\pm$2.1 & 56.2 & 34 \\
{\tiny \object{$\beta$\,Gru}} & {\tiny M5III} & {\tiny 22 42 40 ; -46 53 04} & {\tiny 2006-08-08 08:13:56} & {\tiny \ldots} & 57.2 & 93 \\
{\tiny \object{$\tau^{4}$\,Eri}} & {\tiny M3/M4III} & {\tiny 03 19 31 ; -21 45 28} & {\tiny 2006-08-08 10:09:22} & 10.6$\pm$1 & 62.9 & 66 \\
\hline
\end{tabular}
\end{center}
\end{table*}

This information illustrates the
two main difficulties of these data. First, the time interval between
the observation of \object{$\beta$\,Gruis} and its calibrator is rather long (from
1h30 to 4h30) in comparison to the mean intermediate time dedicated to
the observation of an interferometric calibrator star. The second difficulty concerns the difference in the sky
position of these calibrators compared to that of \object{$\beta$\,Gruis}. The
differences in position are rather large (from 1 to 4h30 for the right
ascension and from 10 to 55$^\circ$ for the declination), which may
also lead to markedly different atmospheric conditions between the 
source and calibration observations. Consequently, the same
difficulties that affect the determination of the calibrated
visibility of \object{$\pi^{1}$\,Gruis} affect the use of unresolved stars
located at some distance on the sky for calibration. However, the
atmospheric condition differences introduce only an amplitude bias on 
the calibrated visibility of the star (see Sect.~\ref{general_description}),

\begin{equation}
\label{visibility_bias}
V_{\rm sou}=\frac{\widetilde{V}_{\rm sou}}{\alpha\widetilde{V}_{\rm cal}},
\end{equation}
where $\alpha$ is a factor related to the atmospheric condition
differences between the source and the calibrator observations. In the
case of a simple uniform-disk intensity distribution like
\object{$\beta$\,Gruis}, the equivalent angular diameter is independent of the
$\alpha$ factor. This is not the case for more complex models that
describe a source like \object{$\pi^{1}$\,Gruis}, which is surrounded by a
circumstellar envelope. Such models use parameters related to the
$\alpha$ factor, which can lead to incorrect interpretation of the
visibility measurements.

\begin{figure}[h!]
\begin{center}
\begin{minipage}[c]{.46\linewidth}
\includegraphics[width=4.cm]{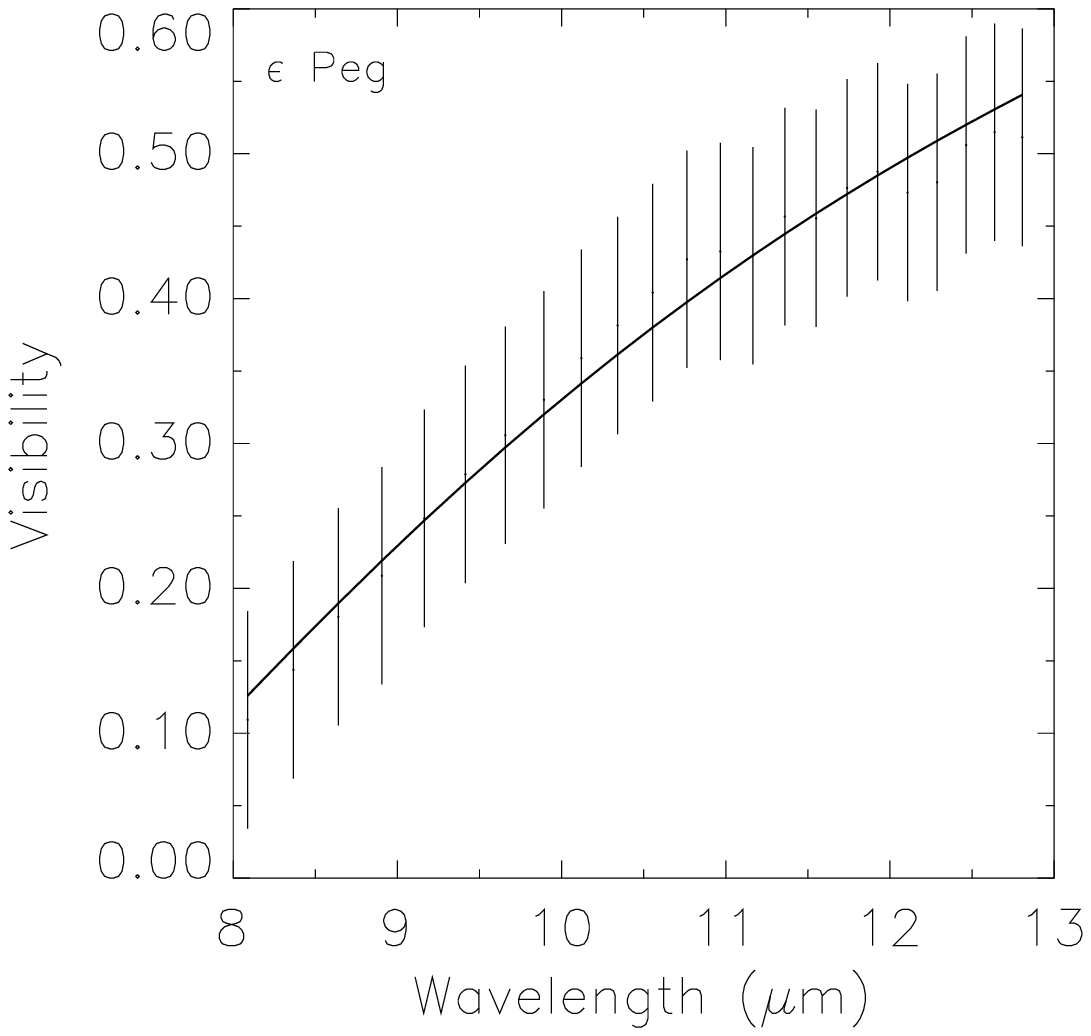}
\end{minipage} \hfill
\begin{minipage}[c]{.46\linewidth}
\includegraphics[width=4.cm]{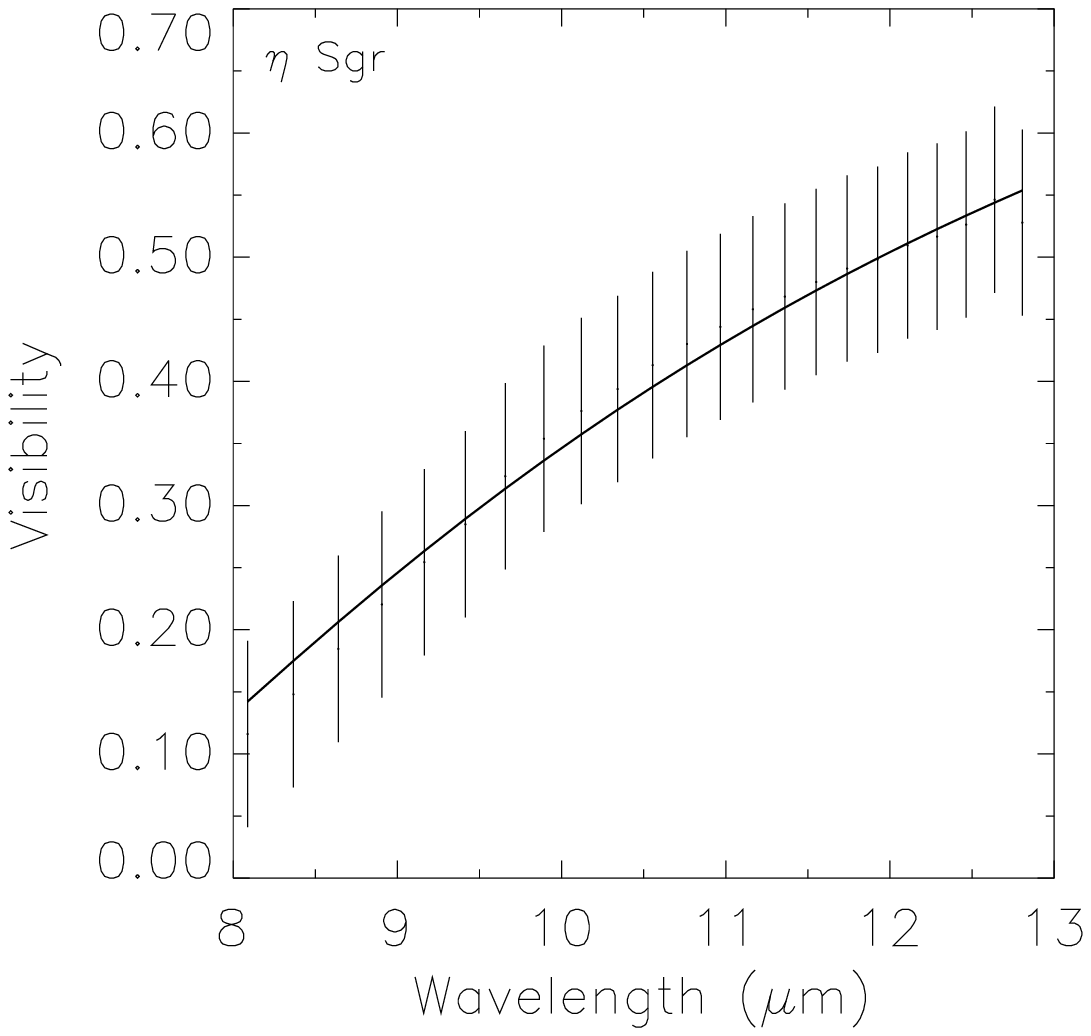}
\end{minipage}
\begin{minipage}[c]{.46\linewidth}
\includegraphics[width=4.cm]{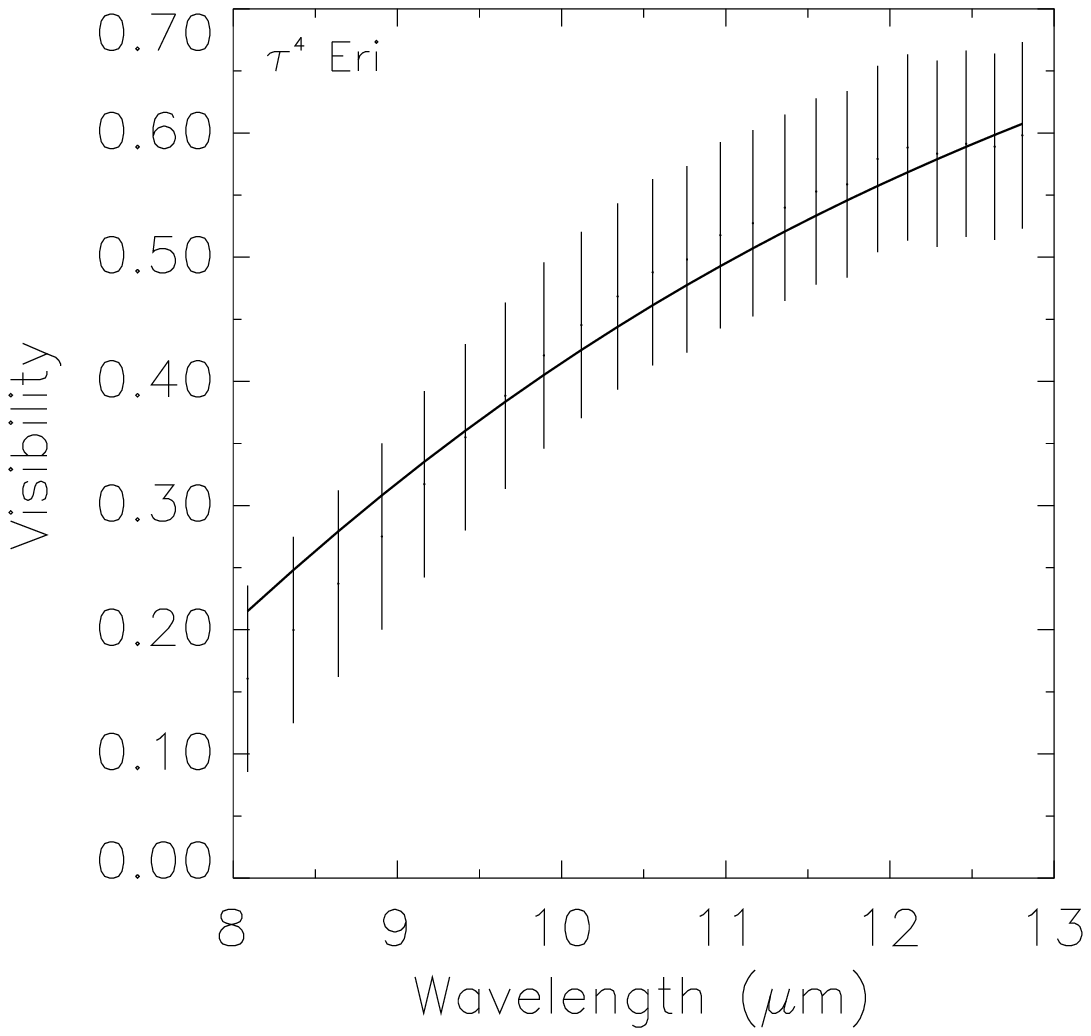}
\end{minipage} \hfill
\begin{minipage}[c]{.46\linewidth}
\caption{Fits of equivalent uniform disks (solid lines) to the calibrated spectrally-dispersed 
visibilities of \object{$\beta$\,Gruis} (error bars) calculated from each of
the 3 calibrators : \object{$\epsilon$\,Peg}, \object{$\eta$\,Sgr}, \object{$\tau^{4}$\,Eri}.}
\end{minipage}
\label{UD_beta}
\end{center}
\end{figure}

Figure~\ref{UD_beta} shows the fits of the equivalent uniform disks to the calibrated
spectrally-dispersed visibilities of \object{$\beta$\,Gruis} calculated from each of the 3 
calibrators (see Table~\ref{journal_2}). The calibrated visibilities
were derived with the MIA and EWS packages and exhibit very good
agreement between the data sets. The error bars on the calibrated
visibilities are typically between 10 and 15\%. 
The mean of the equivalent diameters of the 3 fits is 28.8
and the corresponding error is 0.6 mas.

Table~\ref{beta_diameter} lists the angular diameters of \object{$\beta$\,Gruis}
found in the literature and those derived in this paper using the
spectrometric and interferometric methods.\\

\begin{table}[h!]
\caption{Determination of the angular diameter (in mas) of
\object{$\beta$\,Gruis} with different methods.}
\label{beta_diameter}
\begin{center}
\begin{tabular}{cc} \hline\hline
Reference & Value (mas) \\ \hline
{\tiny \citet{hera02}} & 28.11$\pm$2.82 \\
{\tiny \citet{bedd94}} & 27$\pm$3 \\
{\tiny Spectrometric method (this work)} & 26.8$\pm$1.3 \\
{\tiny Interferometric method (this work)} & 28.8$\pm$0.6 \\
\hline
\end{tabular}
\end{center}
\end{table}

Diameters determined from these last two methods will be taken into account to reduce calibrated visibilities of \object{$\pi^{1}$\,Gruis}. The best set of calibrated visibility data will then be the one giving the
most reliable parameter values, for which we will be able to
simultaneously fit the source spectrum and visibilities.

\listofobjects

\end{document}

%% file: 8306ref.tex
%
%

%
%